\documentclass[useAMS,usenatbib]{mn2e}
\usepackage{graphicx,amssymb,color}
\usepackage[normalem]{ulem}
\newcommand{\rev}{ }

\title[Geometry of WD accretion]
{The entry geometry and velocity of planetary debris into the Roche sphere of a white dwarf}

\author[]{Dimitri Veras$^{1,2}$\thanks{E-mail: d.veras@warwick.ac.uk}\thanks{STFC Ernest Rutherford Fellow},
Nikolaos Georgakarakos$^{3,4}$, 
Alexander J. Mustill$^{5}$, 
\newauthor
Uri Malamud$^{6,7}$,
Tim Cunningham$^{1}$, 
Ian Dobbs-Dixon$^{3,4,8}$
\\
$^{1}$Centre for Exoplanets and Habitability, University of Warwick, Coventry CV4 7AL, UK
\\
$^{2}$Department of Physics, University of Warwick, Coventry CV4 7AL, UK
\\
$^{3}$New York University Abu Dhabi, Saadiyat Island, PO Box 129188, Abu Dhabi, UAE
\\
$^{4}$Center for Astro, Particle and Planetary Physics (CAP$^3$), New York University Abu Dhabi, UAE
\\
$^{5}$Lund Observatory, Department of Astronomy and Theoretical Physics, Lund University, Box 43, SE-221 00 Lund, Sweden
\\
$^{6}$Department of Physics, Technion -- Israel Institute of Technology, Technion City, 3200003 Haifa, Israel
\\
$^{7}$School of the Environment and Earth Sciences, Tel Aviv University, Ramat Aviv, 6997801 Tel Aviv, Israel
\\
$^{8}$Center for Space Sciences, New York University Abu Dhabi, UAE
}

\pubyear{2021}

\begin{document}
\label{firstpage}
\pagerange{\pageref{firstpage}--\pageref{lastpage}}
\maketitle

\begin{abstract}
Our knowledge of white dwarf planetary systems predominately arises from the region within a few Solar radii of the white dwarfs, where minor planets break up, form rings and discs, and accrete onto the star. The entry location, angle and speed into this Roche sphere has rarely been explored but crucially determines the initial geometry of the debris, accretion rates onto the photosphere, and ultimately the composition of the minor planet. Here we evolve a total of over $10^5$ asteroids with single-planet $N$-body simulations across the giant branch and white dwarf stellar evolution phases to quantify the geometry of asteroid injection into the white dwarf Roche sphere as a function of planetary mass and eccentricity. We find that lower planetary masses increase the extent of anisotropic injection and decrease the probability of head-on (normal to the Roche sphere) encounters. Our results suggest that one can use dynamical activity within the Roche sphere to make inferences about the hidden architectures of these planetary systems.
\end{abstract}

\begin{keywords}
Kuiper belt: general – 
minor planets, asteroids: general – 
planets and satellites: dynamical evolution and stability – 
stars: evolution – 
white dwarfs – 
stars: AGB and post-AGB.
\end{keywords}

\section{Introduction}

A near-ubiquitous feature of white dwarf planetary systems is planetary debris near or in the star's photosphere. At least one planetary metal has been detected in over 1,000 white dwarf photospheres, and deep uniform surveys suggest that 25-50 per cent of the white dwarf population contains these metals \citep{vanmaanen1917,vanmaanen1919,dufetal2007,zucetal2010,kleetal2013,koeetal2014,kepetal2015,kepetal2016,couetal2019}. Also, circumstellar dusty and gaseous debris in the form of streams, rings or discs accompanies about 60 of these systems, corresponding to about 1-3 per cent of the population \citep{vanetal2015,farihi2016,manetal2019,manetal2020,vanderboschetal2020}, with more to come \citep{denetal2020,genetal2020,meletal2020,xuetal2020,guietal2021}. 

However, detections of terrestrial or giant planets orbiting white dwarfs have been less frequent, and currently include a handful of only giant planets \citep{thoetal1993,sigetal2003,luhetal2011,ganetal2019,vanetal2020}. This disparity in the abundances of different observational signatures motivates theoretical investigations which attempt to link planetary system architectures with the debris and accretion onto white dwarfs. A goal is to uncover the hidden architectures through our observations of debris and accretion.

Achieving this link is challenging firstly because of the large parameter space to explore, and secondly because of the expensive numerical computations needed both for studying the long-term $N$-body evolution of planetary systems, and for studying the dust and gas dynamics of the rings and discs. As a result, a multitude of theoretical investigations, which are summarized in \cite{veras2016,veras2021}, have focused on specific regions, timescales, or processes in isolation. Indeed, the wide variety of known white dwarf planetary systems and the diversity of planet and stellar multiplicities and architectures in main-sequence systems place doubt on the existence of a one-size-fits-all approach or theory.

Nevertheless, one important physical process which links the observations and theory is the dynamical delivery of debris or larger objects (asteroids, comets, moons, planets) to the immediate vicinity of the white dwarf. By the term ``immediate vicinity" we refer to a few Solar radii, within which most observations take place. Coincidentally, this region is also where tidal disruption \citep{jura2003,jura2008,debetal2012,veretal2014c,rafikov2018,malper2020a,malper2020b} or rotational disruption \citep{veretal2020a} occurs. The extent of this disruption region, at least with respect to tidal disruption, is often referred to as the Roche limit. By assuming that this region is spherical, we equate ``Roche limit" with ``Roche radius". The Roche sphere varies in extent depending on the physical properties (density, spin, shape) of the disrupting object.

Within $N$-body codes, the flagging of a minor planet entering the white dwarf Roche sphere has been a common feature of dynamical investigations \citep{bonetal2011,debetal2012,frehan2014,veretal2014d,bonver2015,hampor2016,payetal2017,petmun2017,steetal2017,musetal2018,smaetal2018,smaetal2021}. However, rarely has the geometry and velocity of entry into the white dwarf Roche sphere, or even the Roche spheres of main-sequence stars \citep{chuetal2020}, been addressed. In one post-main-sequence investigation, \cite{musetal2018} found that in three-planet systems, the orbital inclinations of minor planets which eventually accrete onto the white dwarf appear to be broadly isotropic.

These geometric details are crucial. They determine the type of debris structure formed \citep{malper2020a}, the subsequent evolution of the rings and discs \citep{bocraf2011,rafikov2011a,rafikov2011b,rafgar2012,kenbro2017a,kenbro2017b,mirraf2018,verhen2020,maletal2021,rozetal2021,treetal2021} and, ultimately, the accretion rates onto the photosphere. These rates are then used to reconstruct the chemical composition of the destroyed progenitor \citep[e.g.][]{zucetal2007,kleetal2010,faretal2013,xuetal2017,haretal2018,holetal2018,doyetal2019,swaetal2019}.

Given this strong motivation, the focus of this paper is on quantifying the geometry and velocity of debris which enters the white dwarf Roche sphere. Accruing a sufficiently high number of these encounters to detect trends with planetary parameters requires an extensive suite of $N$-body simulations to be run, primarily because minor planet engulfment into the white dwarf Roche sphere is often less frequent than other types of instability outcomes. 

We perform these simulations here. Given the extent of the simulations, our results also include secondary results of interest, including resonant scattering and escape dynamics. In Section 2, we chronicle in detail our numerical setup. We then describe our results in Section 3 and discuss them in Sections 4-5 before summarizing in Section 6.

\section{Numerical setup}

We aimed to adopt the simplest possible plausible architectures that would generate useful results, helping us to generate a sufficiently high number of realistic encounters between asteroids and the white dwarf Roche sphere. 

Hence, we simulated one-planet systems containing test particles (as proxies for minor planets) around a progenitor host star mass of $2.0M_{\odot}$. This value corresponds to the approximate peak of progenitor mass frequency distribution of the currently observed white dwarf population \citep{treetal2016,cumetal2018,elbetal2018,mccetal2020,barcha2021}. We used the {\tt SSE} code from \cite{huretal2000} to model a solar metallicity star of this mass. This code evolves the star for 1.16~Gyr along the main-sequence and about 330~Myr along the giant branch phases, such that the star reaches the white dwarf phase with only about 32 per cent of its original mass.

As for the physical radius of the white dwarf, that value is only about 1 per cent of its Roche sphere, and the extent of the Roche sphere is dependent on the physical properties of the minor planet. In order to obtain a self-consistent high-resolution set of results, we fixed the Roche sphere for all test particles at $1R_{\odot}$. This value roughly corresponds to a spinning rocky minor planet around a fiducial white dwarf of mass $0.60-0.65M_{\odot}$ \citep{veretal2017}. Upon entering this Roche sphere, the test particle is removed from the system. 

At the opposite end of the distance spectrum, as the star evolves and loses mass, its Hill surface -- an ellipsoid which defines the escape boundary of the system -- decreases as the 1/3rd power of the stellar mass. We define the Hill surface by assuming that our planetary systems reside approximately 8 kpc away from the Galactic centre, and use the tidal prescriptions from \cite{vereva2013} and \cite{veretal2014a}. Throughout the evolution, the three axes of the Hill ellipsoid are on the order of $10^5$ au. Test particles exceeding this boundary are flagged as escapees; the planet's orbit is fixed, except for being expanded by a factor of about 3.1 due to giant branch mass loss \citep{omarov1962,hadjidemetriou1963,veretal2011}, and never escapes.

We sampled five different planetary masses ($M=1,5,20,100,300M_{\oplus}$) and three initial eccentricities ($e=0.0, 0.2, 0.5$) but kept all other planetary orbital elements fixed at $0^{\circ}$ except for semimajor axis, whose initial value was set at $10.5$ au. This value was chosen in order to sample secular and resonant interactions with interior minor planets over a wide range of resonances, out to and exceeding the $6$:$1$ mean motion resonance.

In each simulation, we included about $10^4$ test particles along with the planet. These test particles were randomly sampled in uniform distributions of semimajor axis between 3 and 10 au and eccentricity between 0.0 and 0.7 such that their initial pericentres lay between 3 and 10 au. Sampled test particles whose orbital pericentres resided outside of this range were not included in the simulation. The lower pericentre bound of 3 au was chosen because it represents the approximate maximum radius of the progenitor asymptotic giant branch star \citep{musvil2012}. 

The inclination of the test particles were randomly sampled from a uniform distribution ranging from $0^{\circ}$ to $5^{\circ}$. This non-coplanarity with the planet is natural (as evidenced by the Solar system) and particularly important: if all of the test particles were coplanar, then not only would the rate of collisions with the planet be artificially high, but also the encounters with the white dwarf Roche sphere could occur only in two dimensions. All of the other orbital elements of the minor planets were randomly sampled from uniform distributions encompassing their entire ranges.

For our simulations, we adopted the planetary evolution code presented in \cite{musetal2018}, within which is incorporated a stellar evolution profile from the {\tt SSE} stellar evolution code \citep{huretal2000}. We used a RADAU integrator with a tolerance of $10^{-12}$ to propagate the orbital timesteps. At and within each of these timesteps, we interpolated the stellar mass and radius, which are provided by {\tt SSE} over much longer timesteps.  

In order to maximize our computational resources, we started the simulations at the beginning of the red giant branch phase, and continued them for 1 Gyr after the star had become a white dwarf. Throughout the white dwarf phase, when a test particle encountered the white dwarf Roche sphere, we recorded the position and velocity of the minor planet at the start of that timestep.

The code does not model radiative effects. Along the giant branch phases these effects are important for minor planets under about $10^3$ km in radius \citep{bonwya2010,donetal2010,veretal2014b,veretal2019,versch2020,zotver2020} but are also untenable to numerically simulate in their full generality \citep{veretal2015a}. Therefore, when considering the test particles in the context of self-consistent evolution, they must be treated as larger than $10^3$ km. Even without this restriction of self-consistency, our dynamical results during the white dwarf phase of evolution are still valid for any size of minor planet or debris. Nevertheless, for ease of language, henceforth we refer to the minor planets (or test particles) as asteroids.

\begin{figure*}
\centerline{
\includegraphics[width=8.5cm]{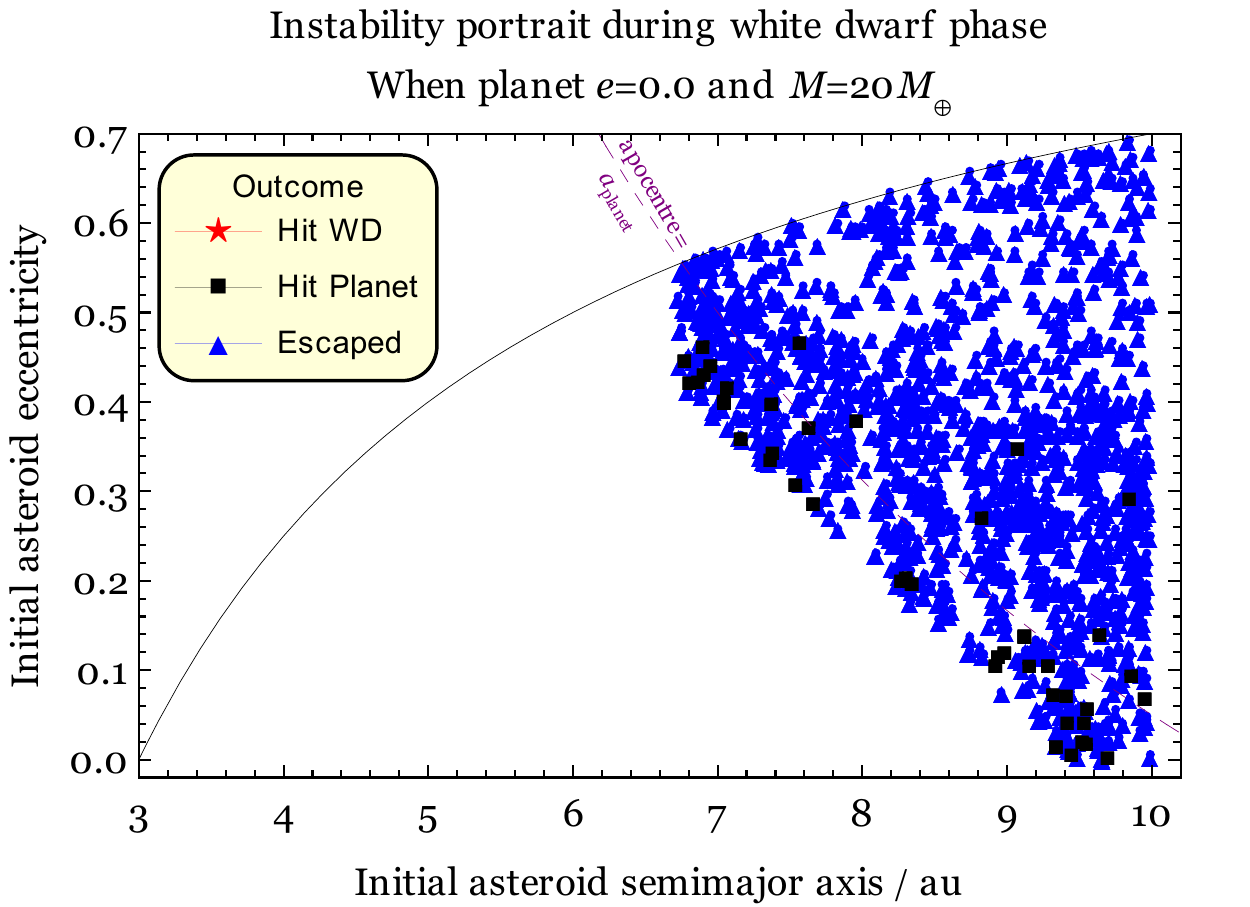}
\includegraphics[width=8.5cm]{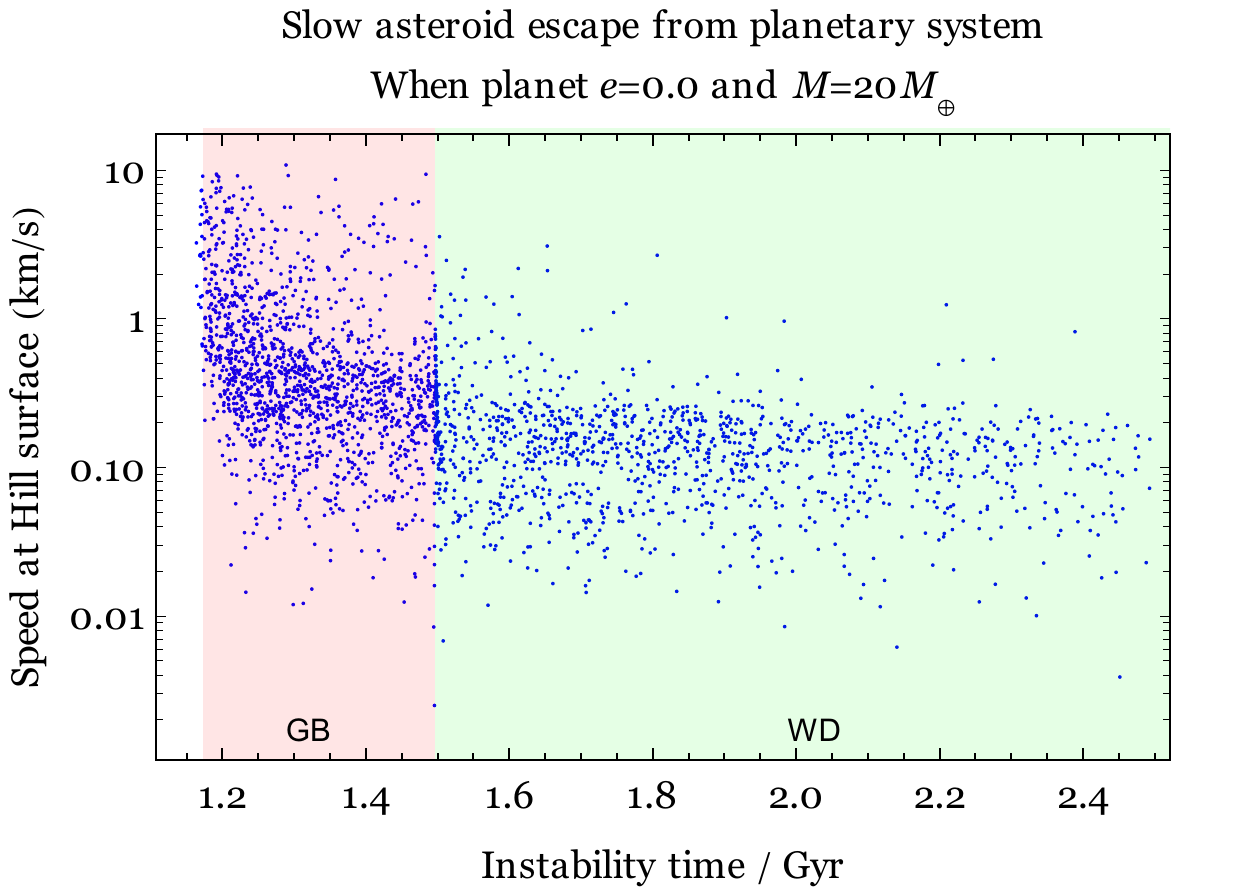}
}
\centerline{}
\centerline{
\includegraphics[width=8.5cm]{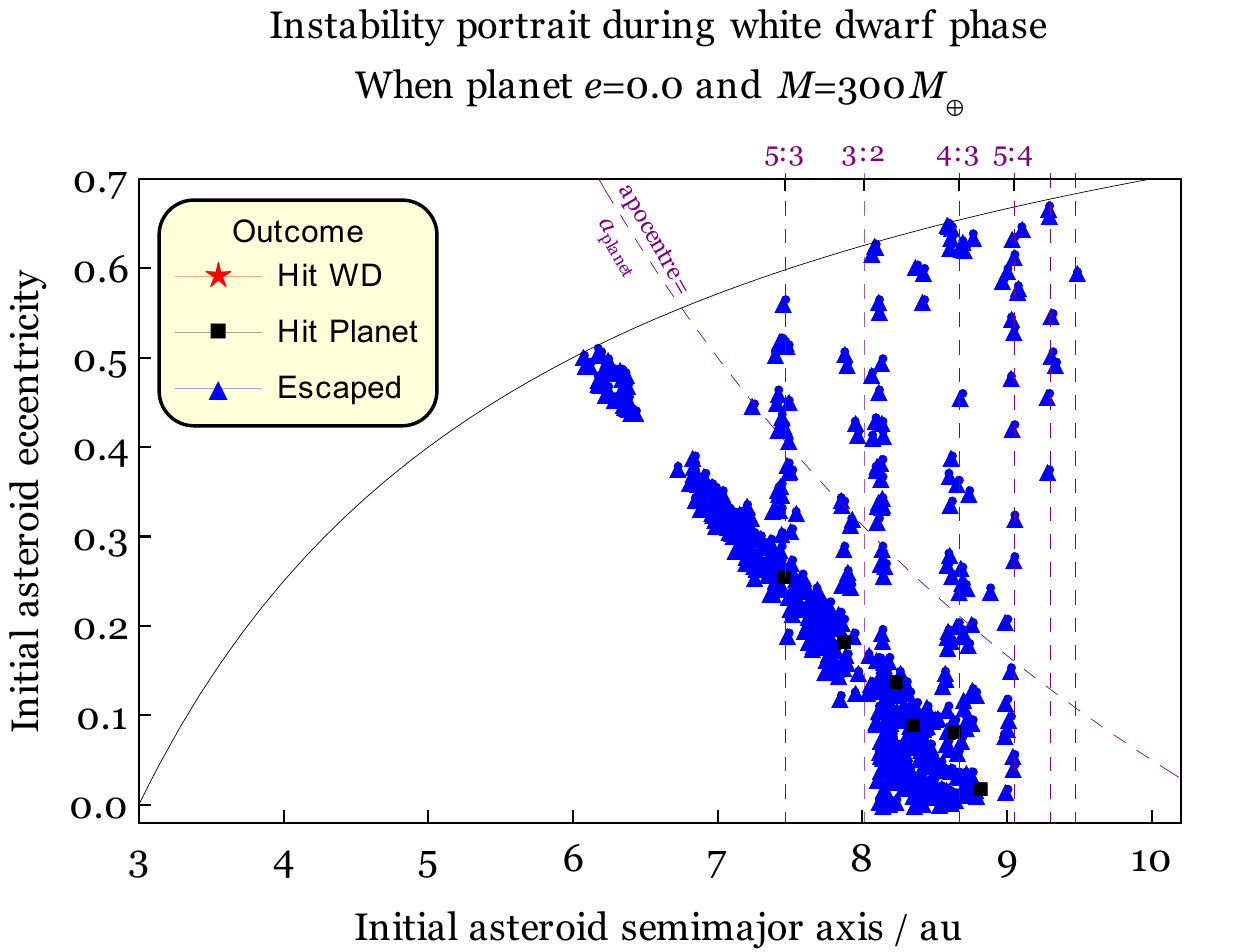}
\includegraphics[width=8.5cm]{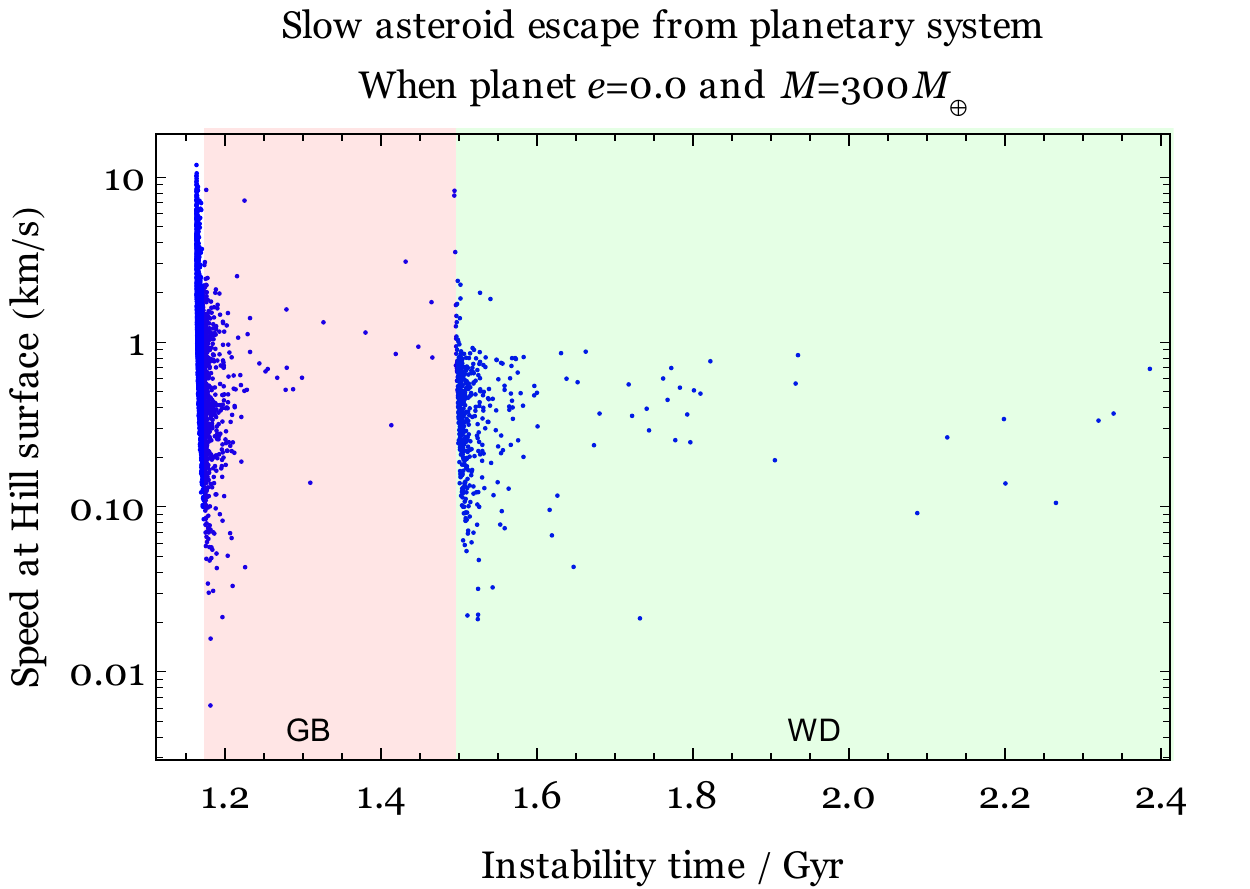}
}
\caption{
Outcomes for simulations where the planet was on a circular orbit. Not shown are objects which remained stable throughout the simulations. {\it Left panels}: All instabilities along the white dwarf phase only. These are entirely in the form of escape from the system and collision with the planet. Resonant features (the {\rev purple} number ratios and {\rev purple} dashed vertical lines) are more prominent for the higher planetary mass. The curved black line represents the boundary in parameter space below which asteroids were sampled, and the curved {\rev purple} dashed line is the boundary where initial asteroid apocentres equalled that of the planet. {\it Right panels}: All instabilities in the simulations. The escape speeds are predominantly sub-km/s, and for the higher planetary mass, the instability timescales are sharply peaked just after the initiation of the simulation and around the peak of giant branch mass loss.
}
\label{e00}
\end{figure*}

\begin{figure*}
{\Huge Instability portraits: low-mass planets}
\centerline{
\includegraphics[width=8.5cm]{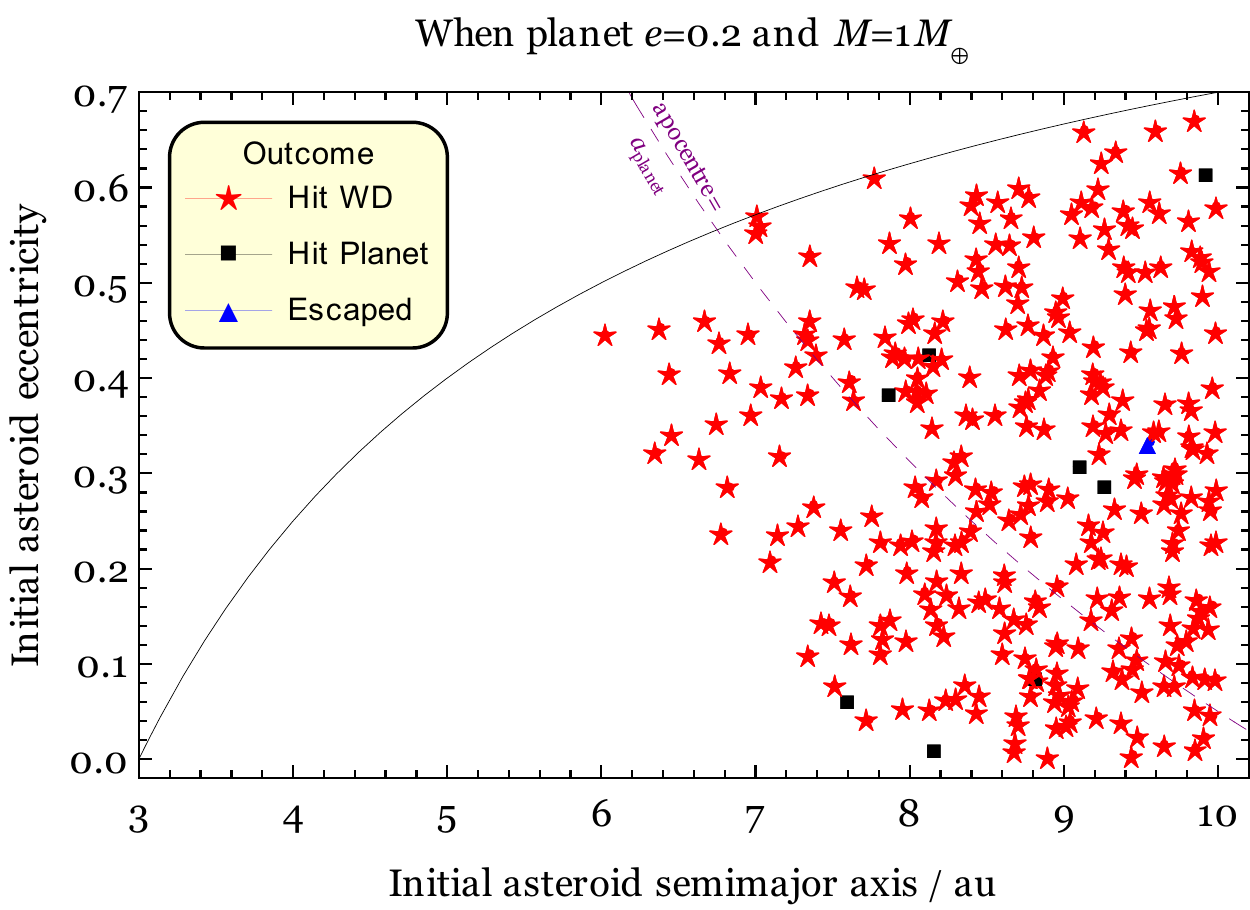}
\includegraphics[width=8.5cm]{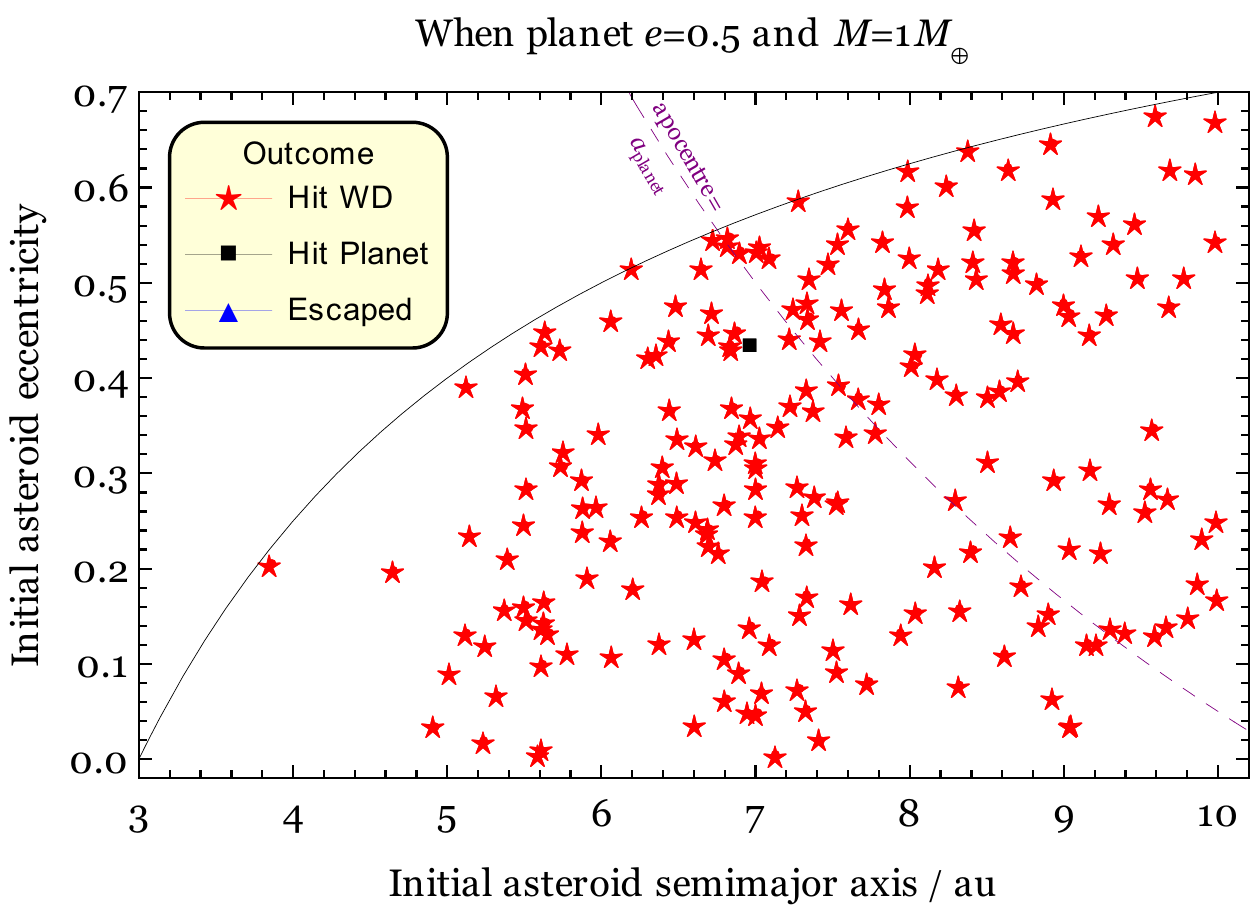}
}
\centerline{
\includegraphics[width=8.5cm]{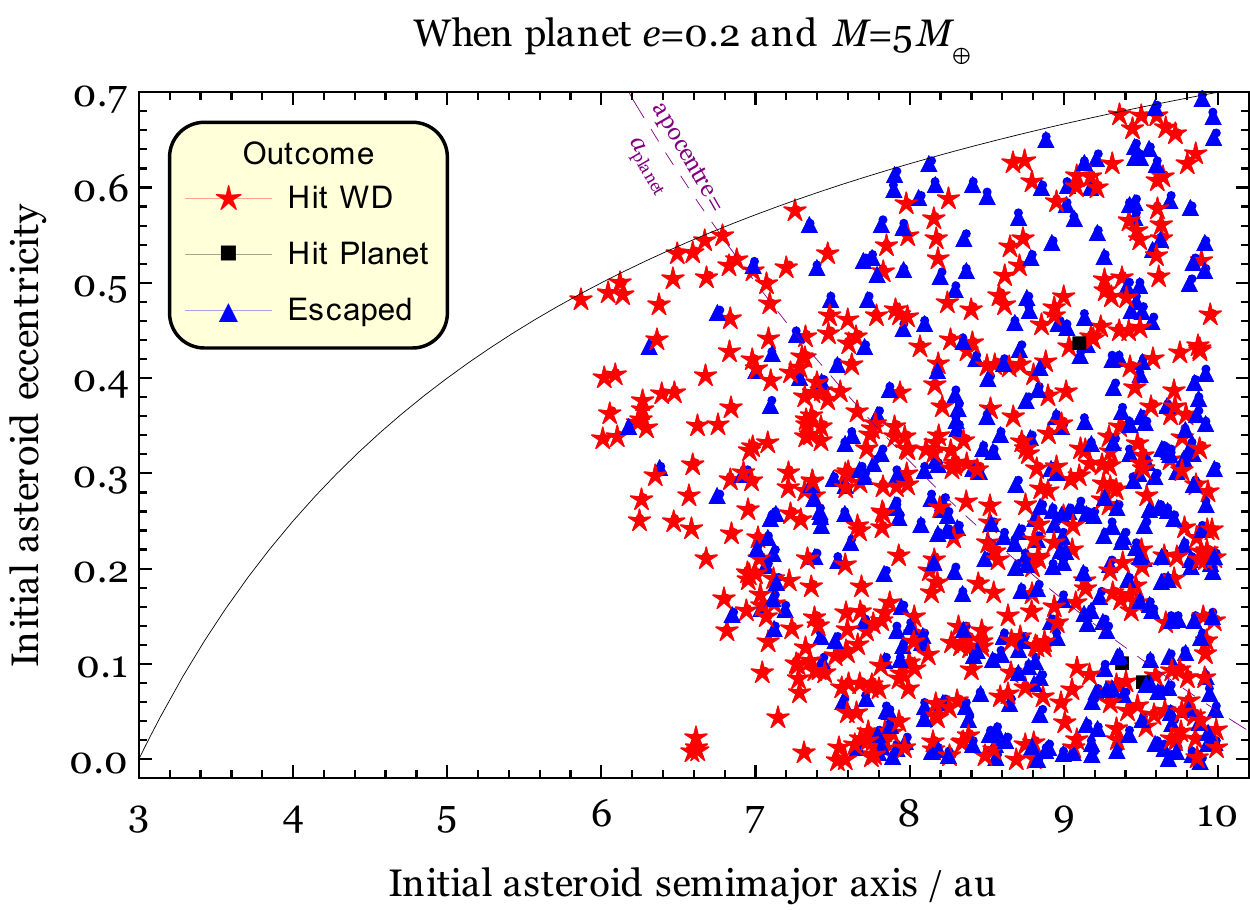}
\includegraphics[width=8.5cm]{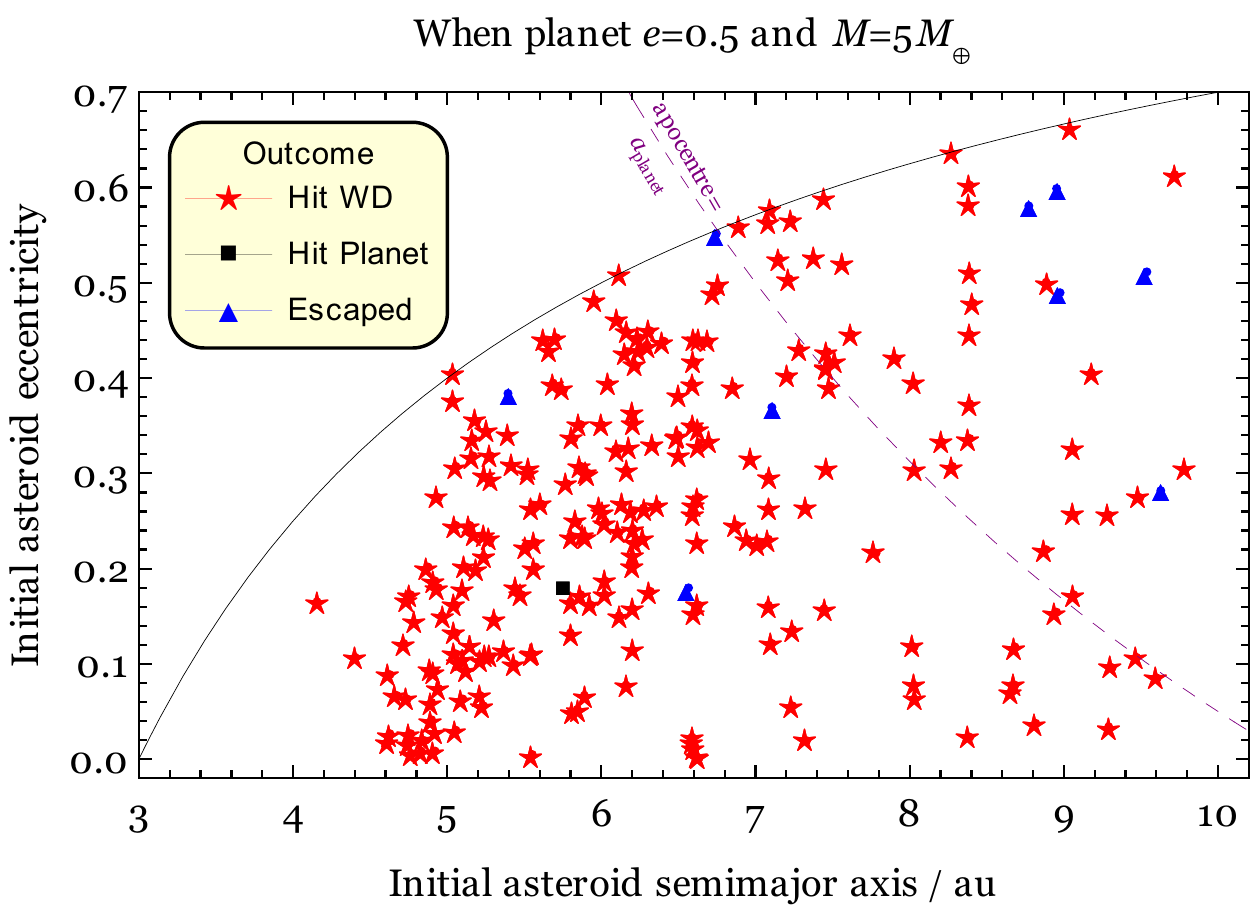}
}
\centerline{}
\centerline{
\includegraphics[width=8.5cm]{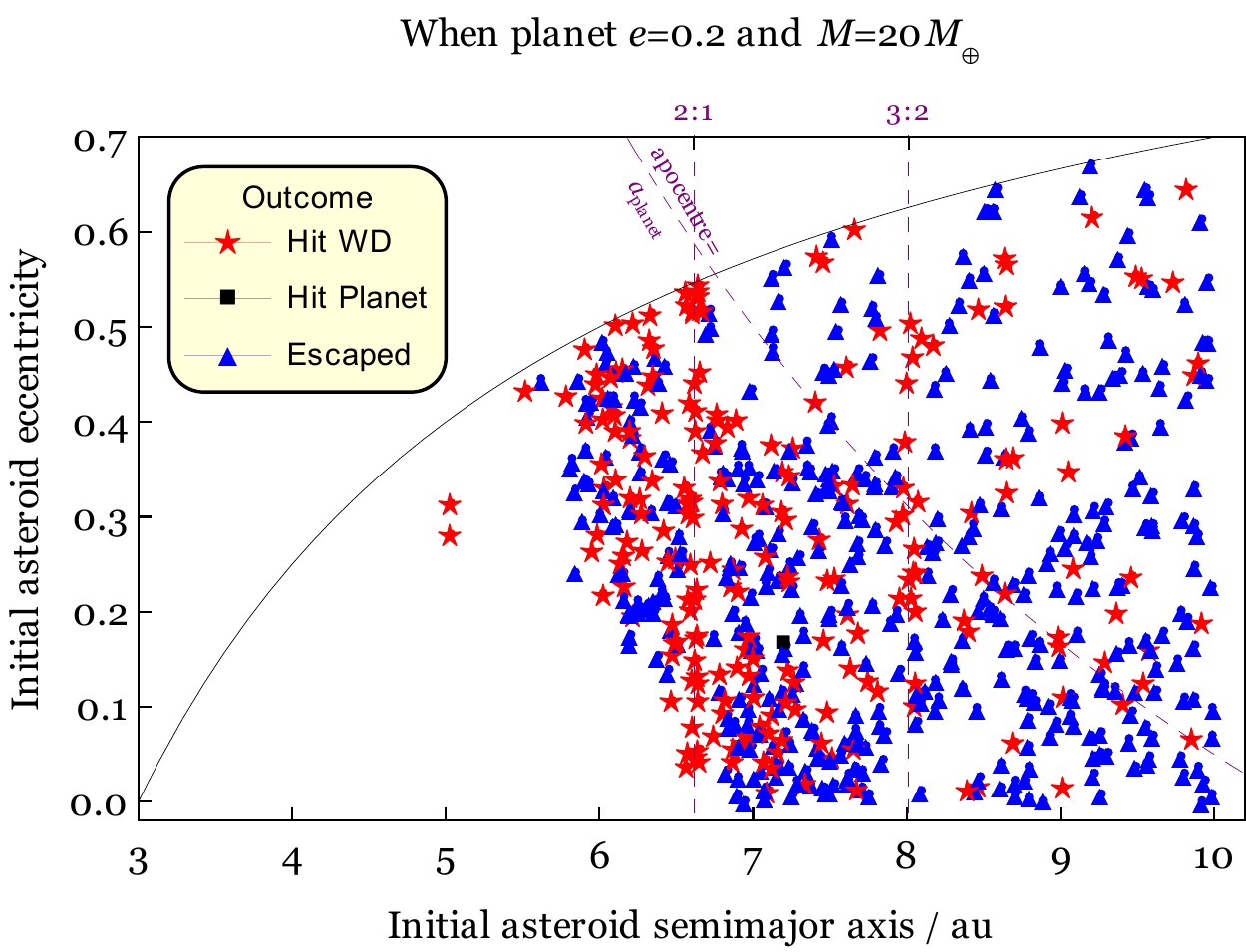}
\includegraphics[width=8.5cm]{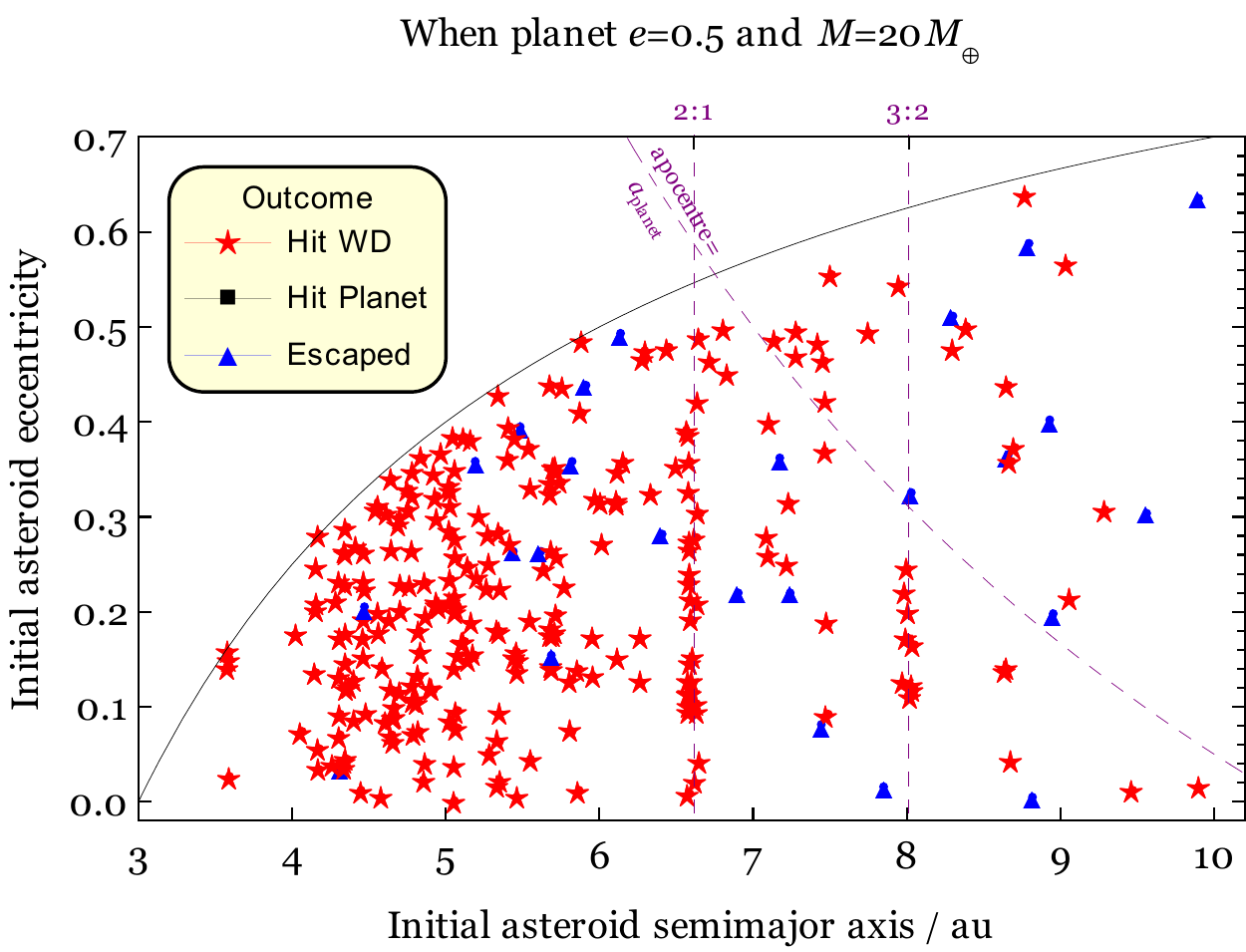}
}
\caption{
Instability portraits along only the white dwarf phase for the simulations with low planetary masses. The left and right columns respectively illustrate the $e=0.2$ and $e=0.5$ cases. The label ``Hit WD" refers to encounters between an asteroid and the white dwarf Roche sphere. Increasing planetary eccentricity allows for more asteroids which are initially close to the white dwarf to be engulfed. For the low planetary masses shown here, instabilities do not predominantly occur at the strongest mean motion resonances {\rev except perhaps for the $M=20M_{\oplus}$ cases, where a transition occurs to the higher mass cases}.
}
\label{inst1}
\end{figure*}

\begin{figure*}
{\Huge Instability portraits: high-mass planets}
\centerline{
\includegraphics[width=8.5cm]{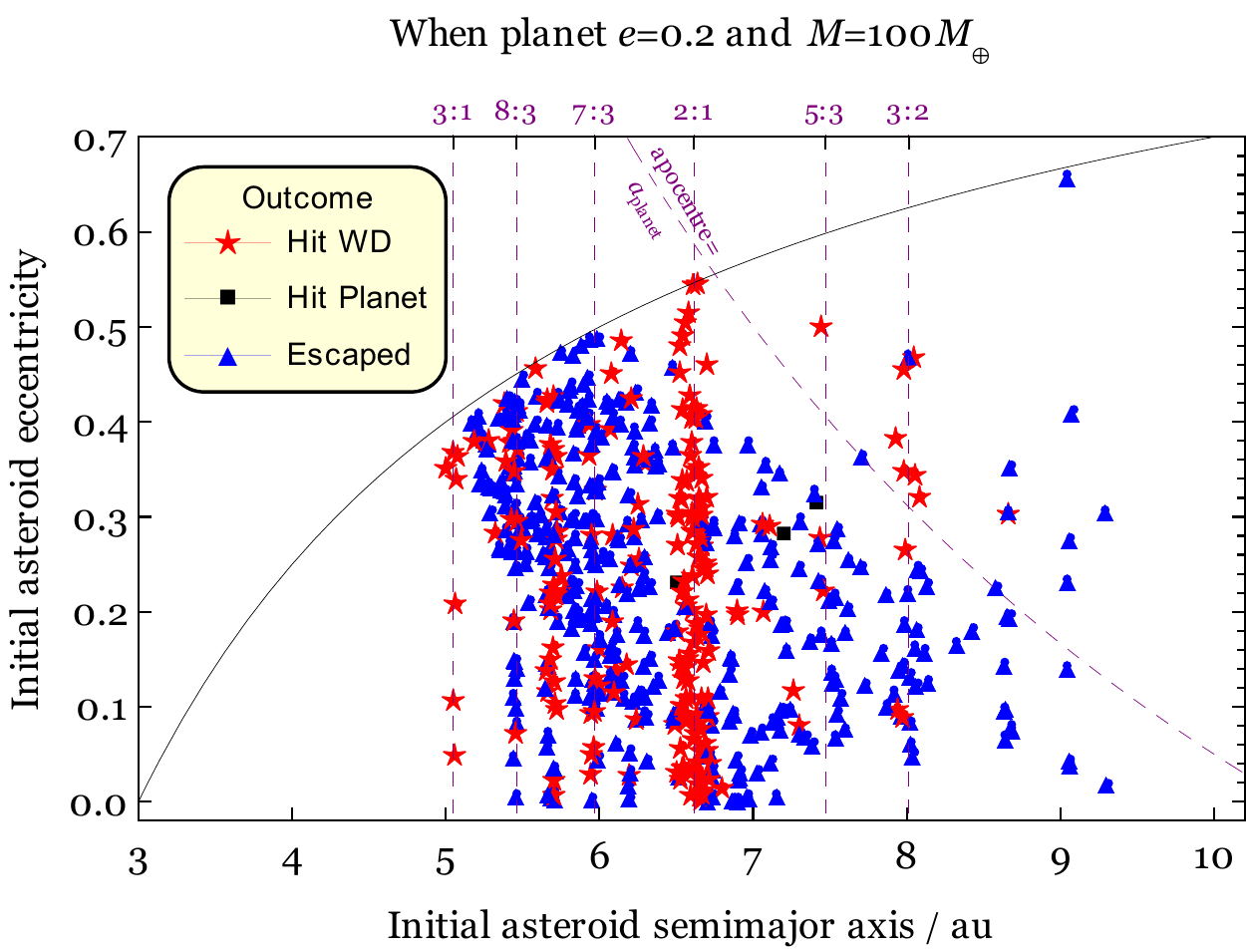}
\includegraphics[width=8.5cm]{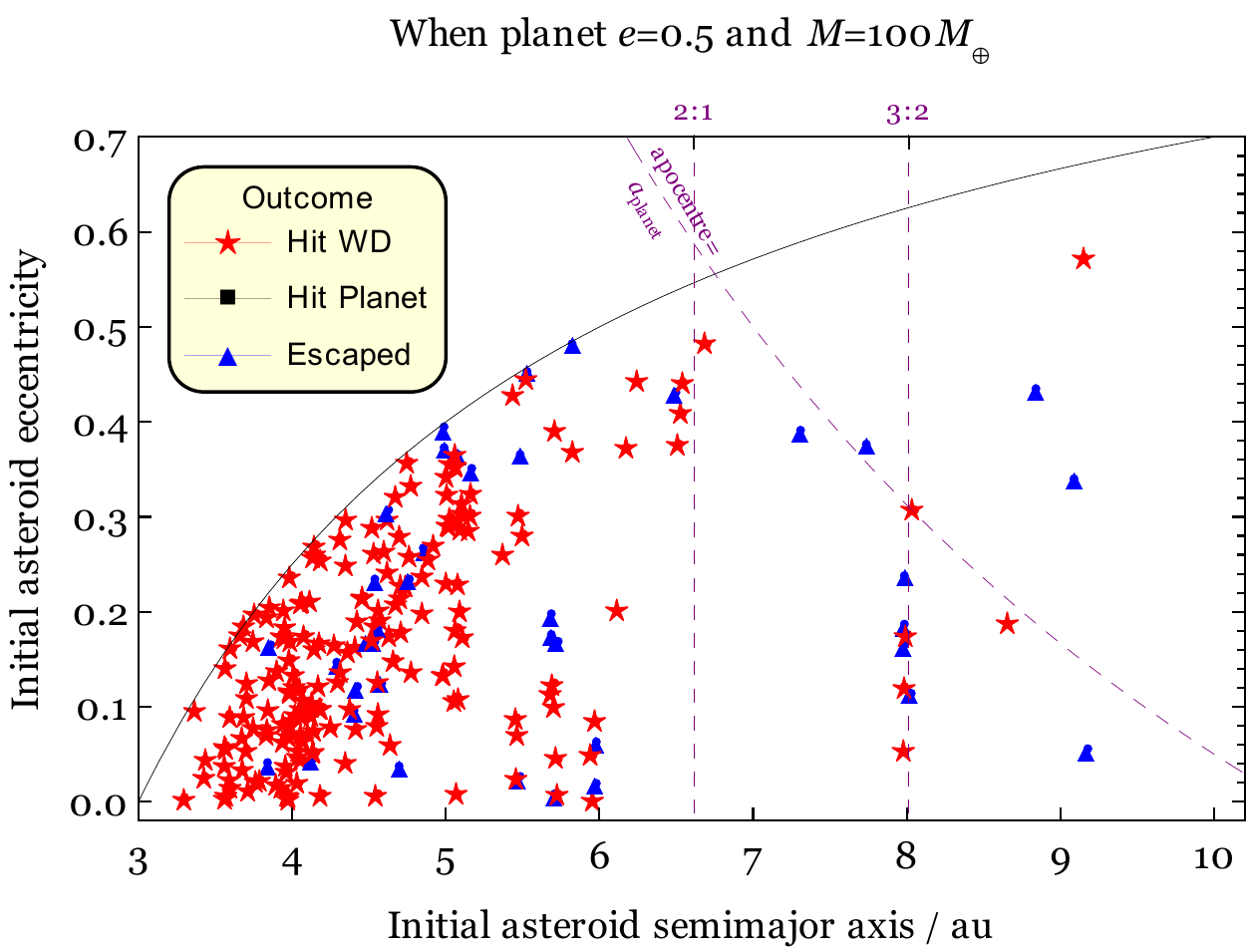}
}
\centerline{}
\centerline{
\includegraphics[width=8.5cm]{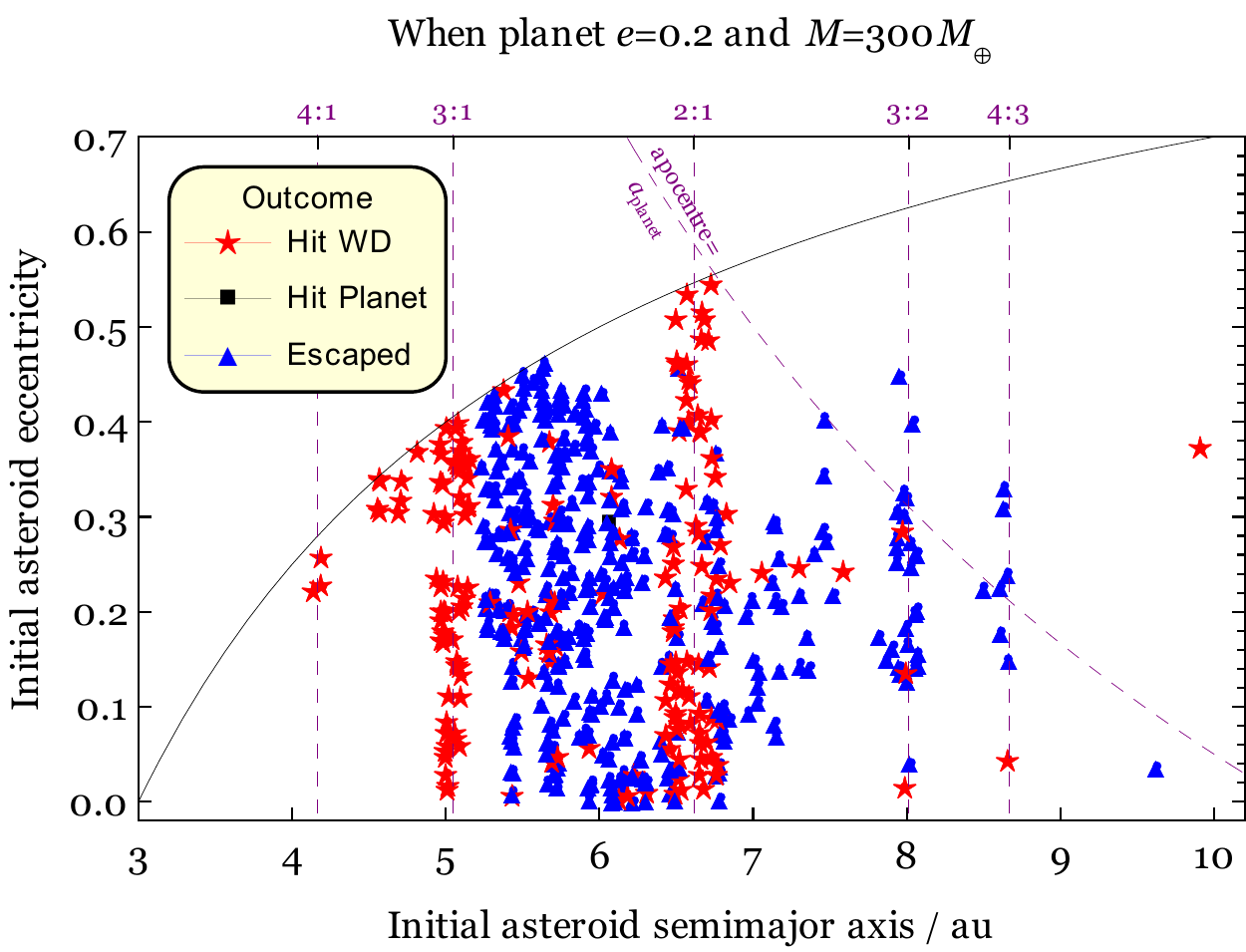}
\includegraphics[width=8.5cm]{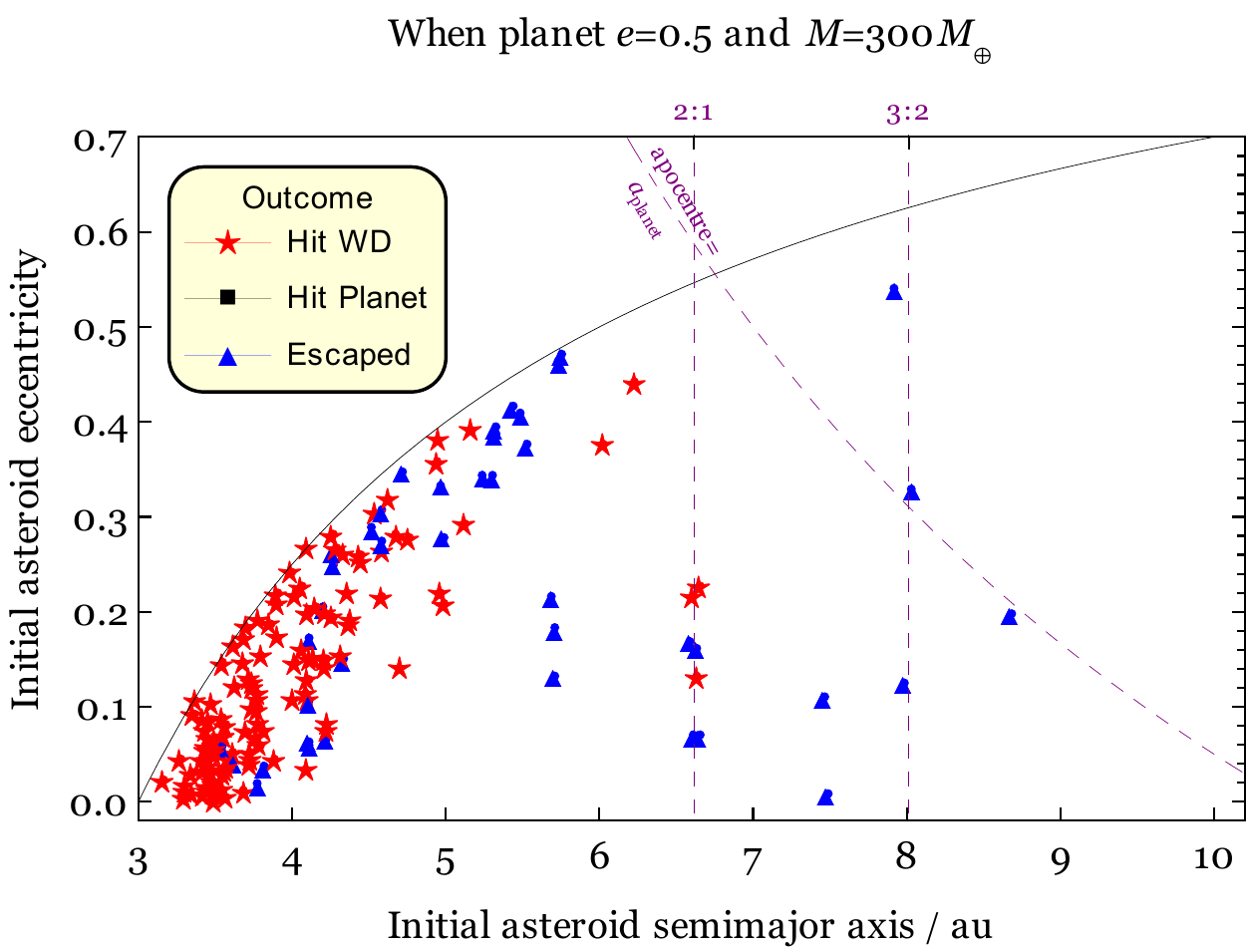}
}
\caption{
Instability portraits along only the white dwarf phase for simulations with high planetary masses. The influence of mean motion resonances is significant for the $e=0.2$ cases, but becomes less distinct for the $e=0.5$ cases. Few instabilities during the white dwarf phase occur for asteroids whose initial apocentre exceeded that of the planet. In the highest mass, highest eccentricity case, nearly all engulfments are due to asteroids with the minimum initial semimajor axis allowed.
}
\label{inst2}
\end{figure*}

\begin{figure}
\includegraphics[width=7.5cm]{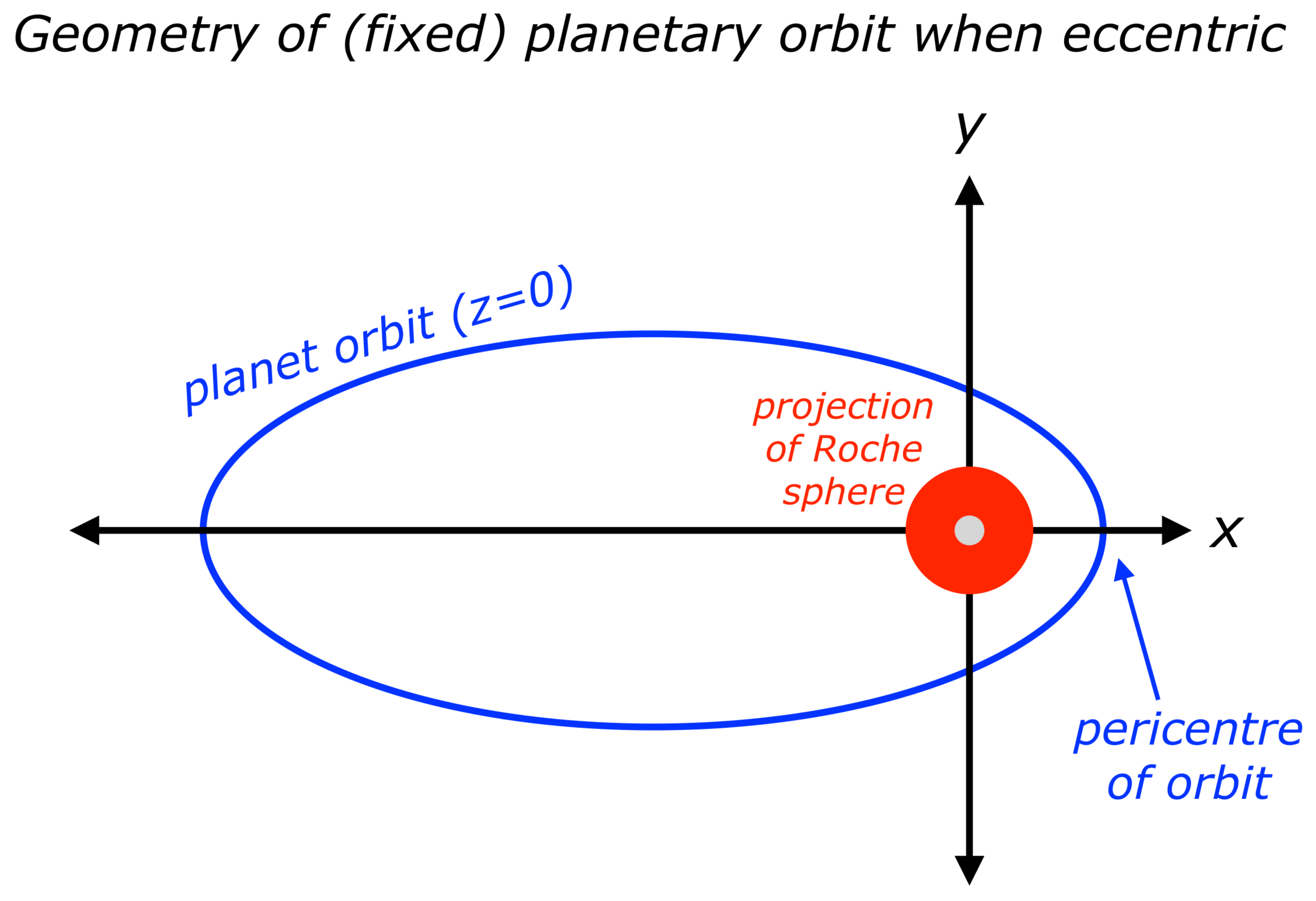}
\caption{
Cartesian geometry of the fixed planetary orbit, which is important for understanding the anisotropy of debris injection into the Roche sphere. The planet travels in a counterclockwise direction.
}
\label{PlanGeom}
\end{figure}

\begin{figure*}
\includegraphics[width=14cm]{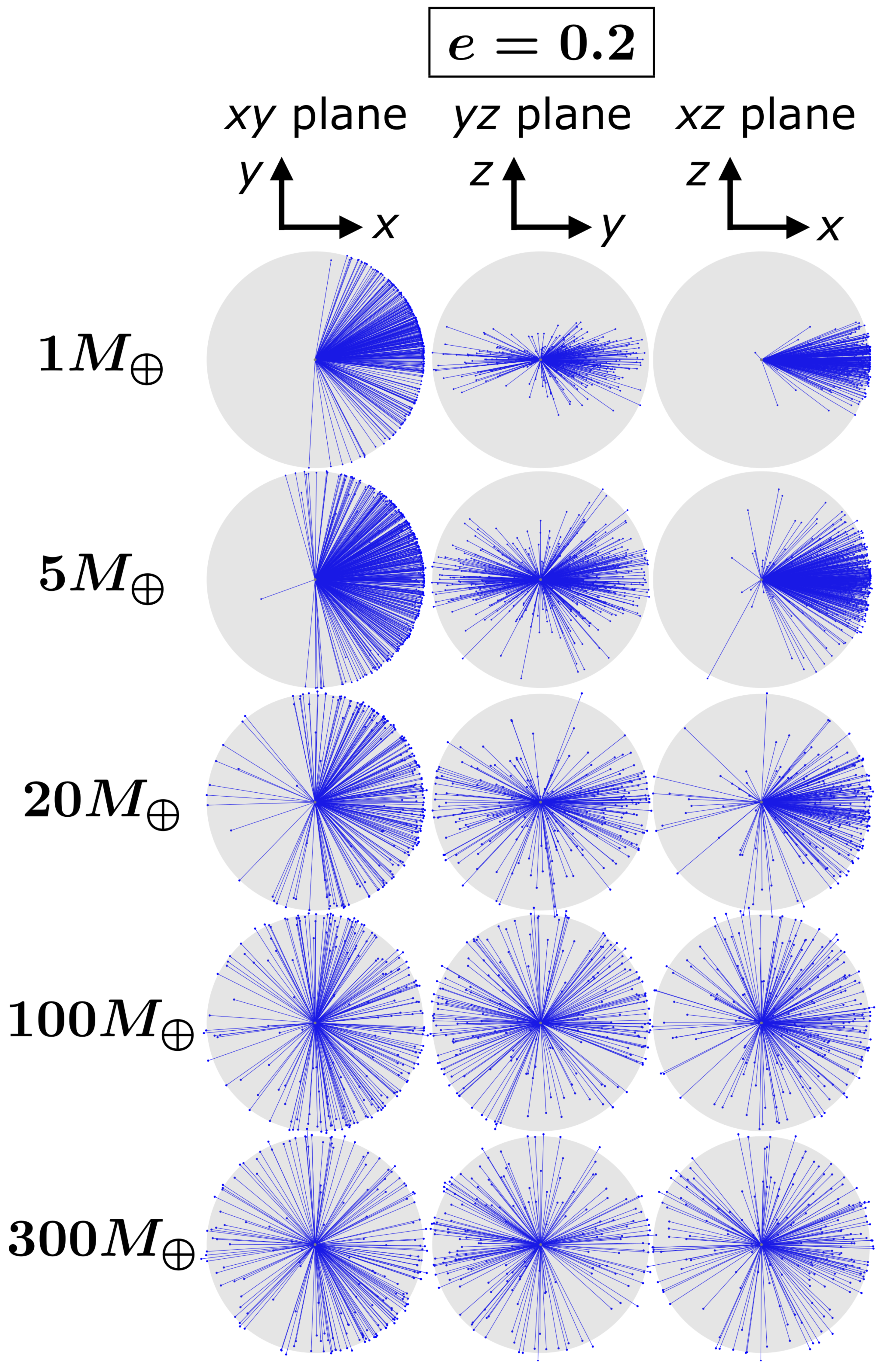}
\caption{
Geometry of asteroid encounters with the white dwarf Roche sphere for $e=0.2$. The projected Roche spheres onto the three Cartesian planes are shown as gray discs, and the locations of the asteroids at the last timestep before entering are shown as dots. These dots are connected with lines to the white dwarf for visual effect. This collage illustrates, for the lowest mass planets, {\rev that} the strong anisotropy of encounters {\rev is} due to {\rev the geometry of the planet's orbit}. As the planet mass increases, the tidal encounter geometry becomes increasingly isotropic.
}
\label{e02P234}
\end{figure*}

\begin{figure*}
\includegraphics[width=14cm]{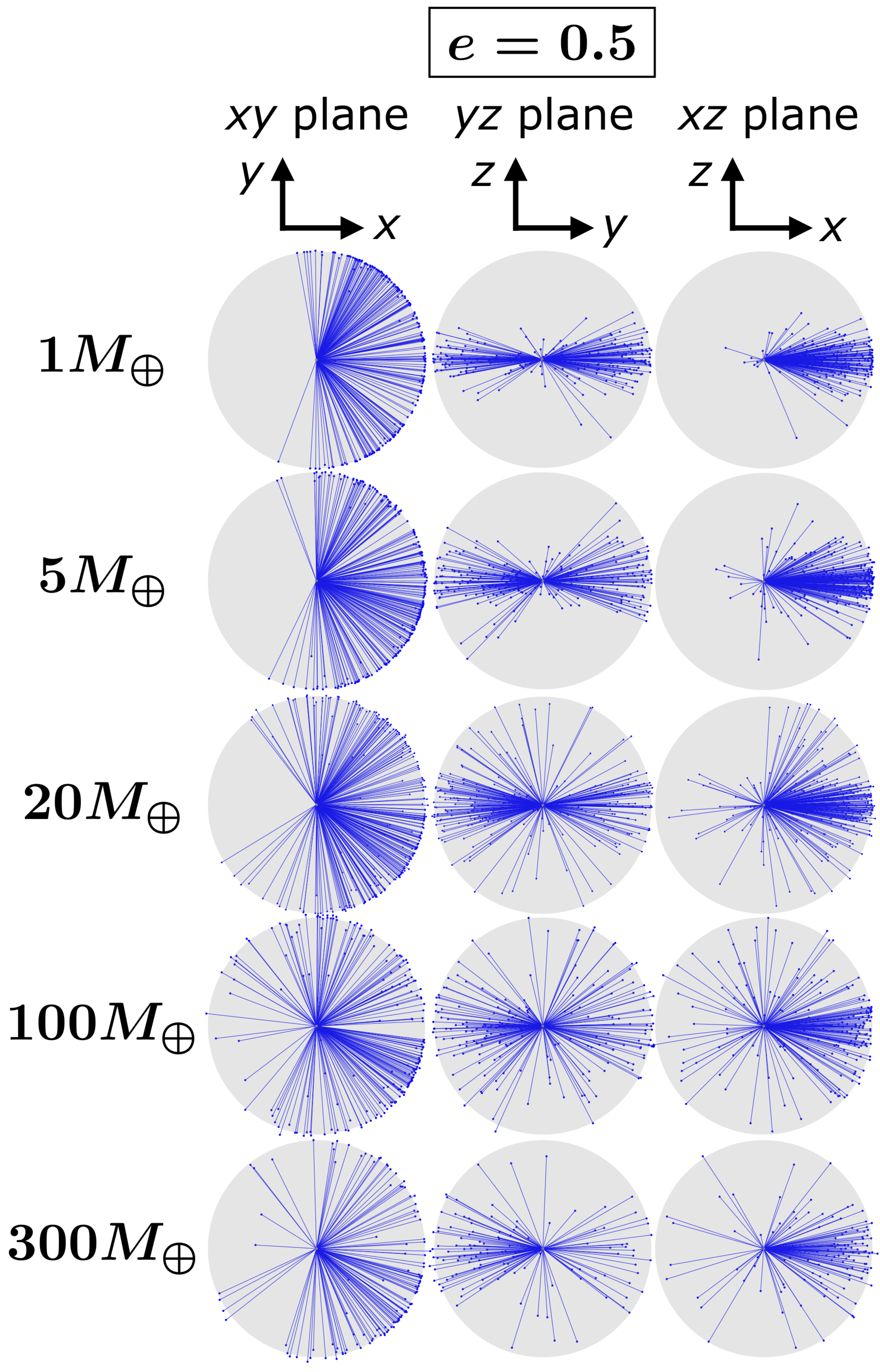}
\caption{
Same as Fig. \ref{e02P234}, but for $e=0.5$. Although the trends in both figures are similar, here the isotropy correlation with planet mass is weaker.
}
\label{e05P234}
\end{figure*}

\begin{figure*}
{\Huge Osculating elements immediately before Roche sphere engulfment}
\centerline{}
\centerline{
\includegraphics[width=8.5cm]{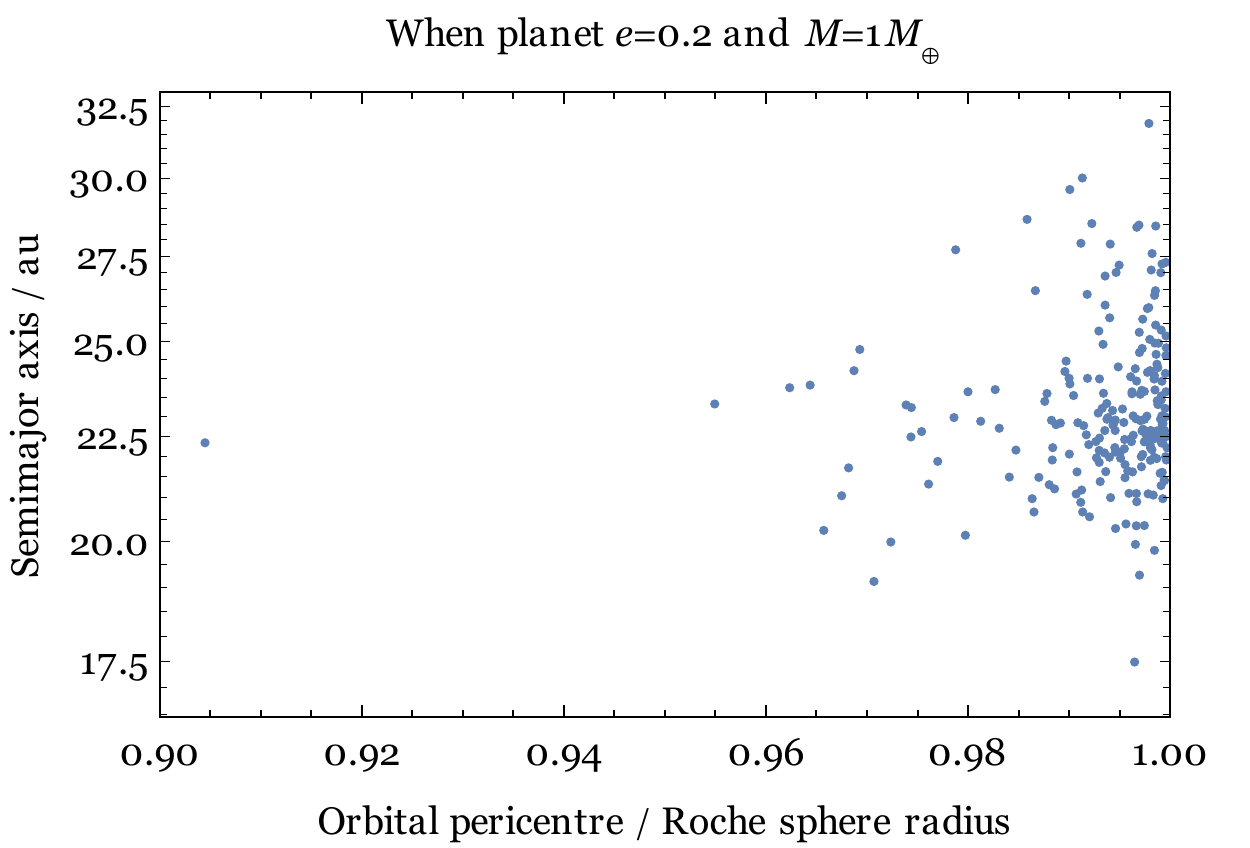}
\includegraphics[width=8.5cm]{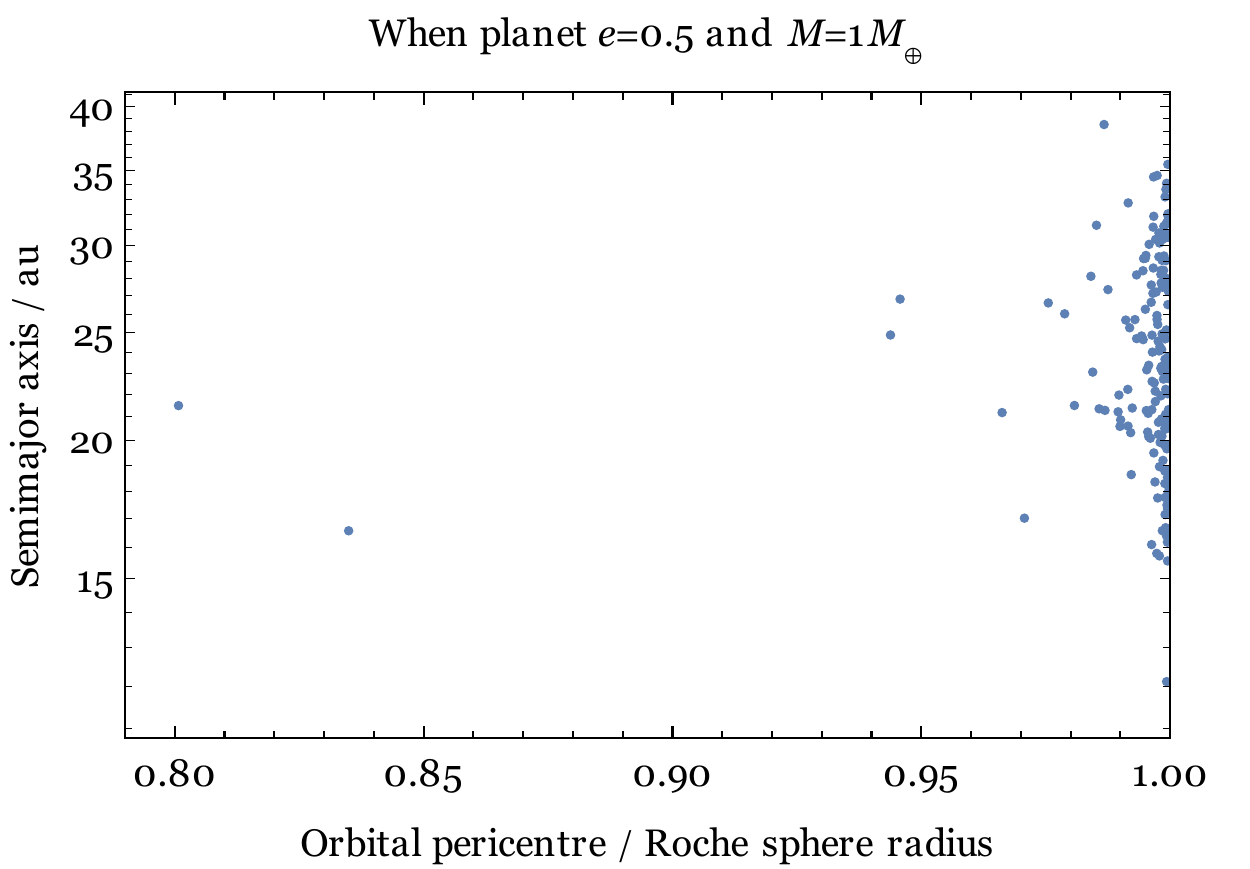}
}
\centerline{}
\centerline{
\includegraphics[width=8.5cm]{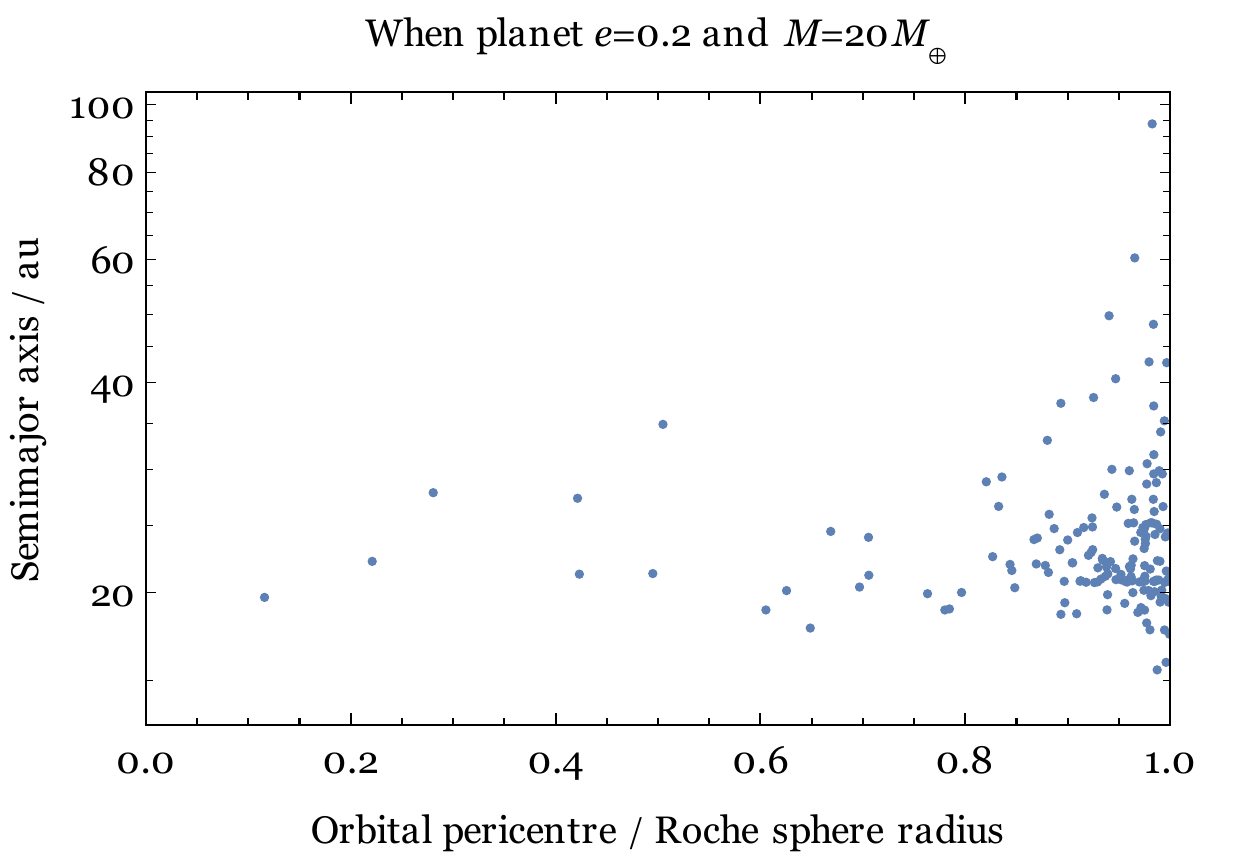}
\includegraphics[width=8.5cm]{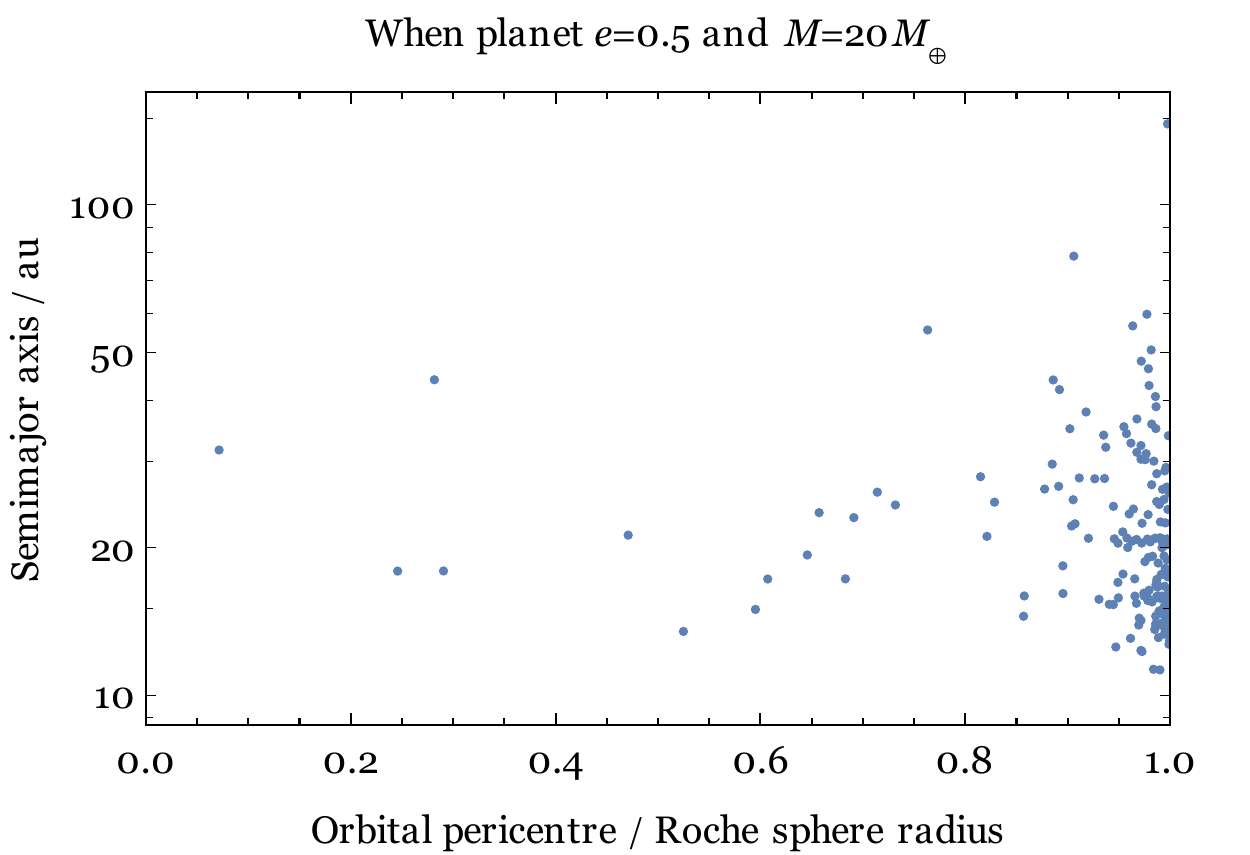}
}
\centerline{}
\centerline{
\includegraphics[width=8.5cm]{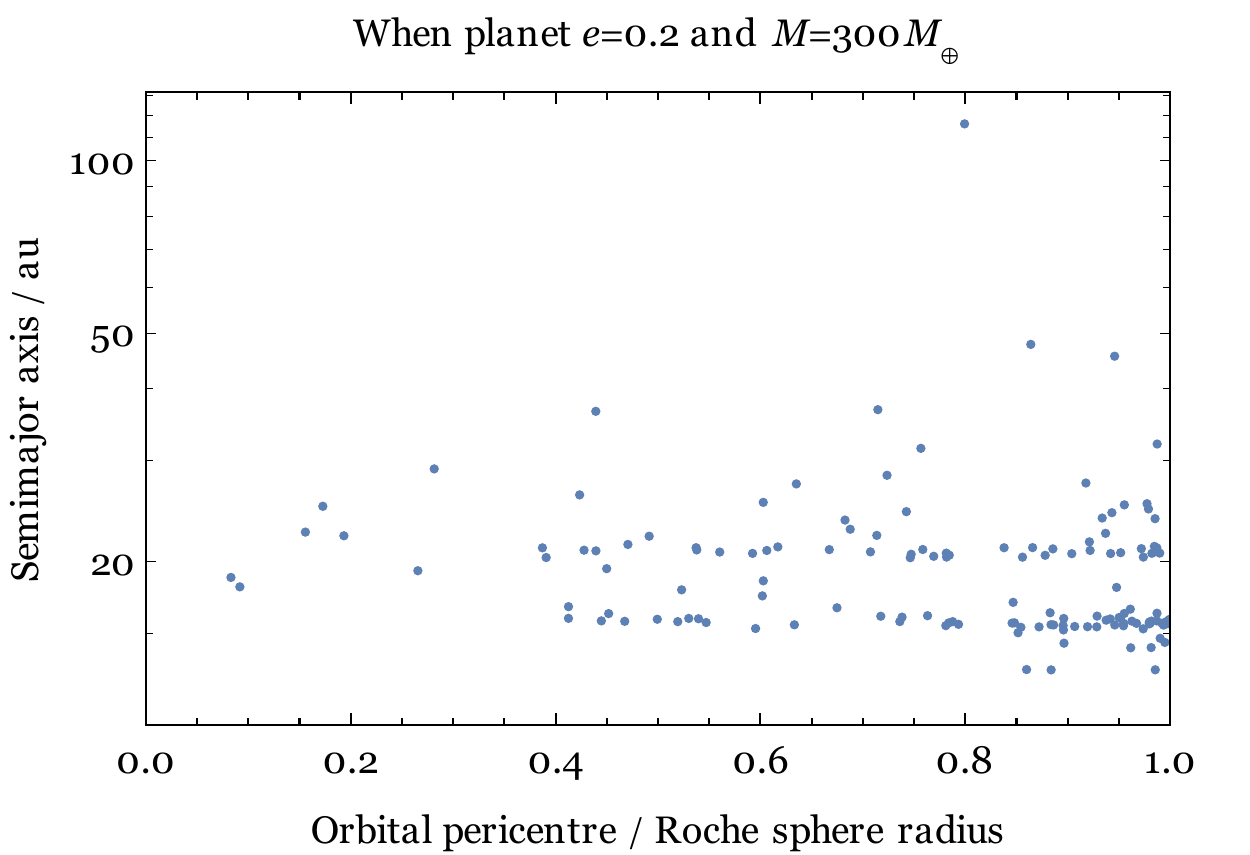}
\includegraphics[width=8.5cm]{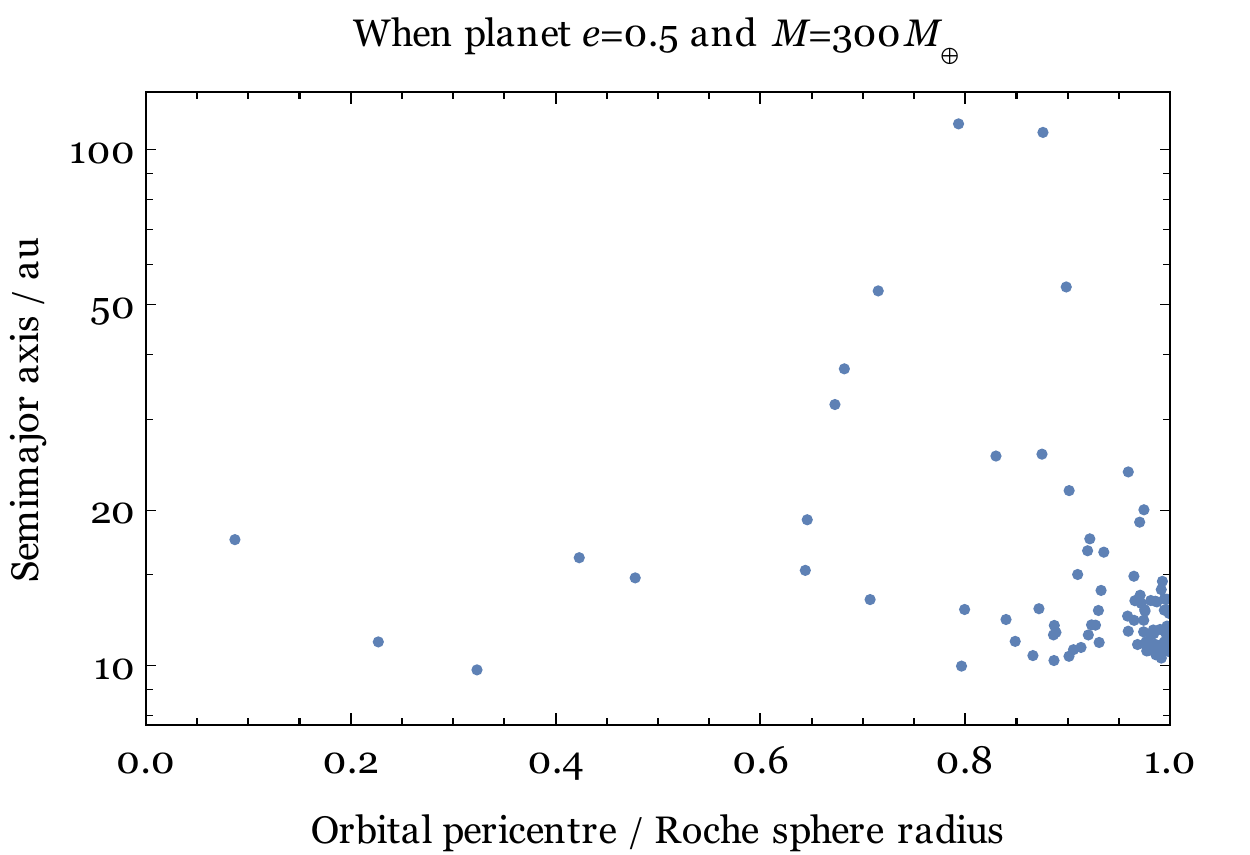}
}
\caption{
The entry point into the white dwarf Roche sphere. Shown are the osculating semimajor axes as a function of their osculating orbital pericentres at the timestep before engulfment. The smaller the pericentre, the more ``head-on" the injection is. The horizontal features in the lower left panel correspond to resonant locations where a significant population of asteroids are perturbed towards the white dwarf (see Fig. \ref{inst2}).
}
\label{roche1}
\end{figure*}

\begin{figure*}
{\Huge Entry speed into the Roche sphere}
\centerline{}
\centerline{
\includegraphics[width=8.5cm]{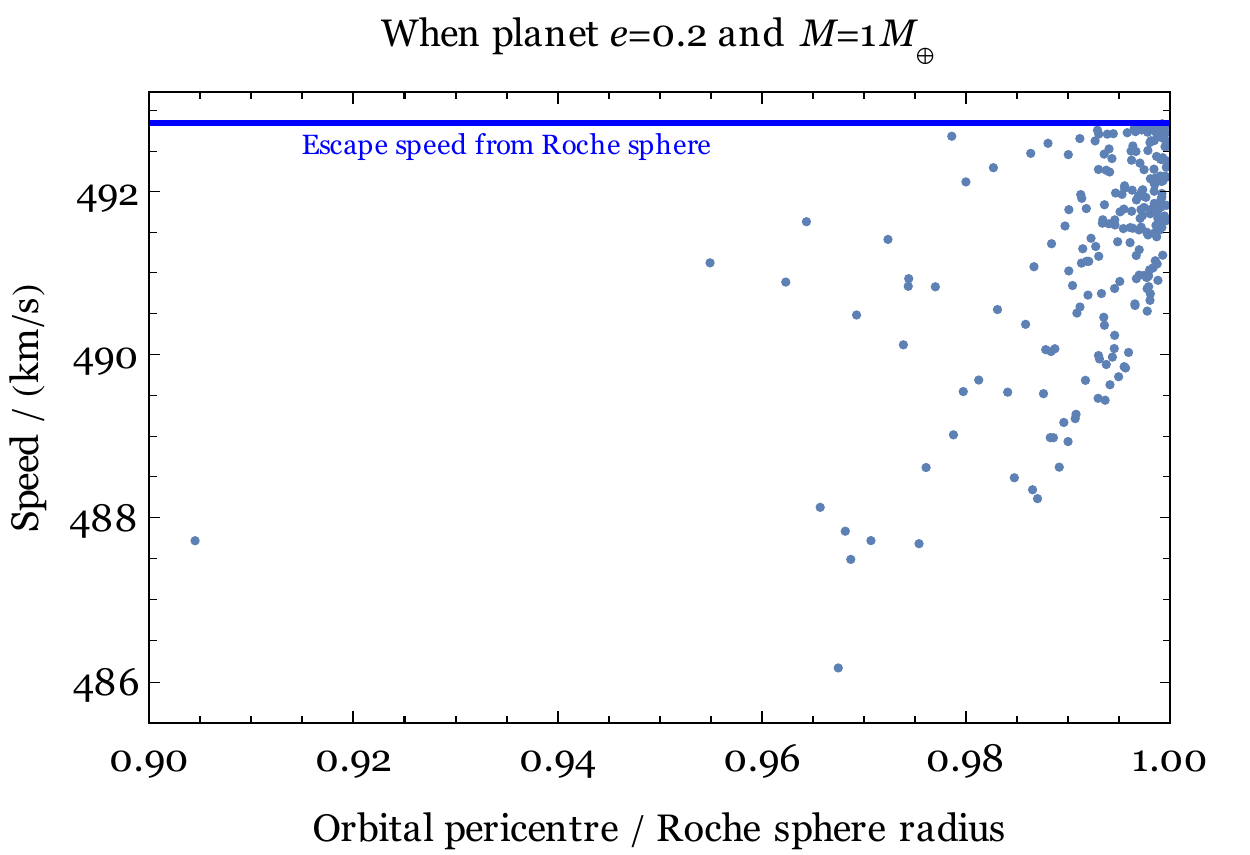}
\includegraphics[width=8.5cm]{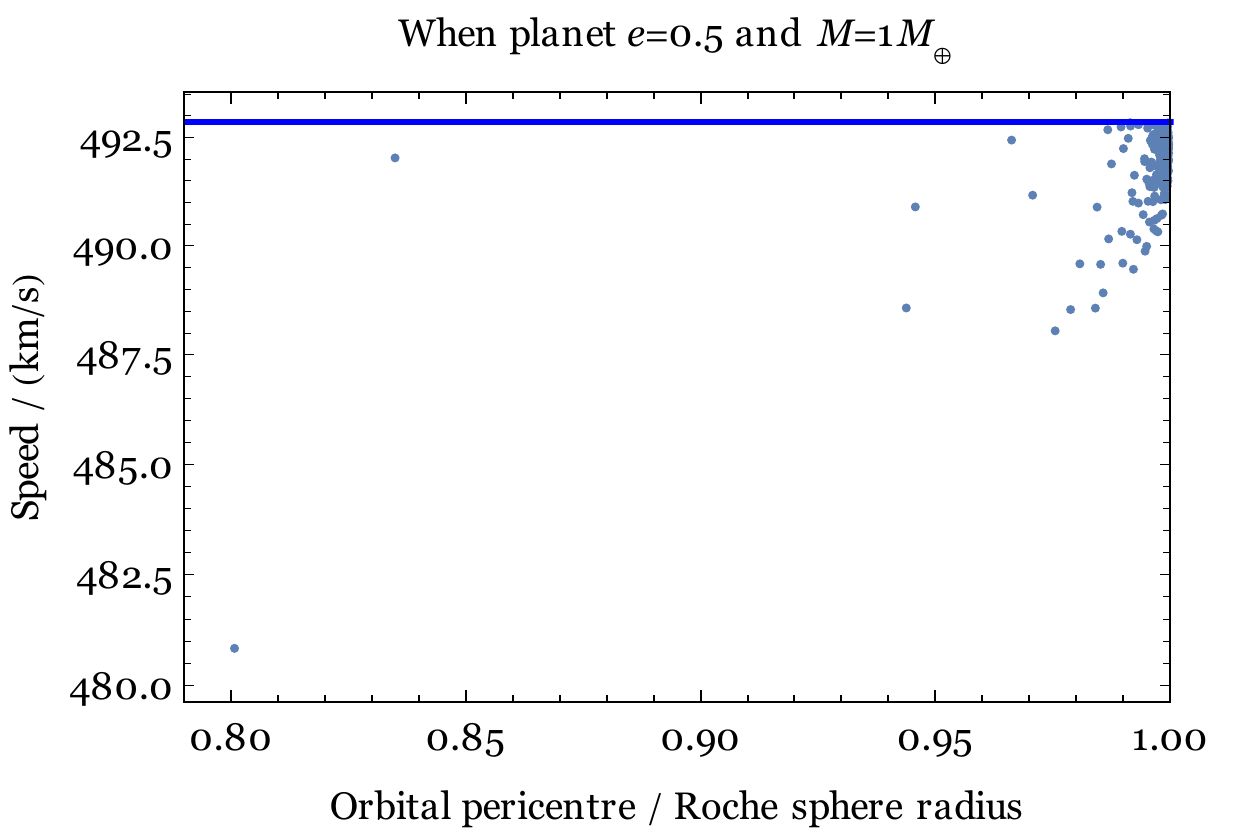}
}
\centerline{}
\centerline{
\includegraphics[width=8.5cm]{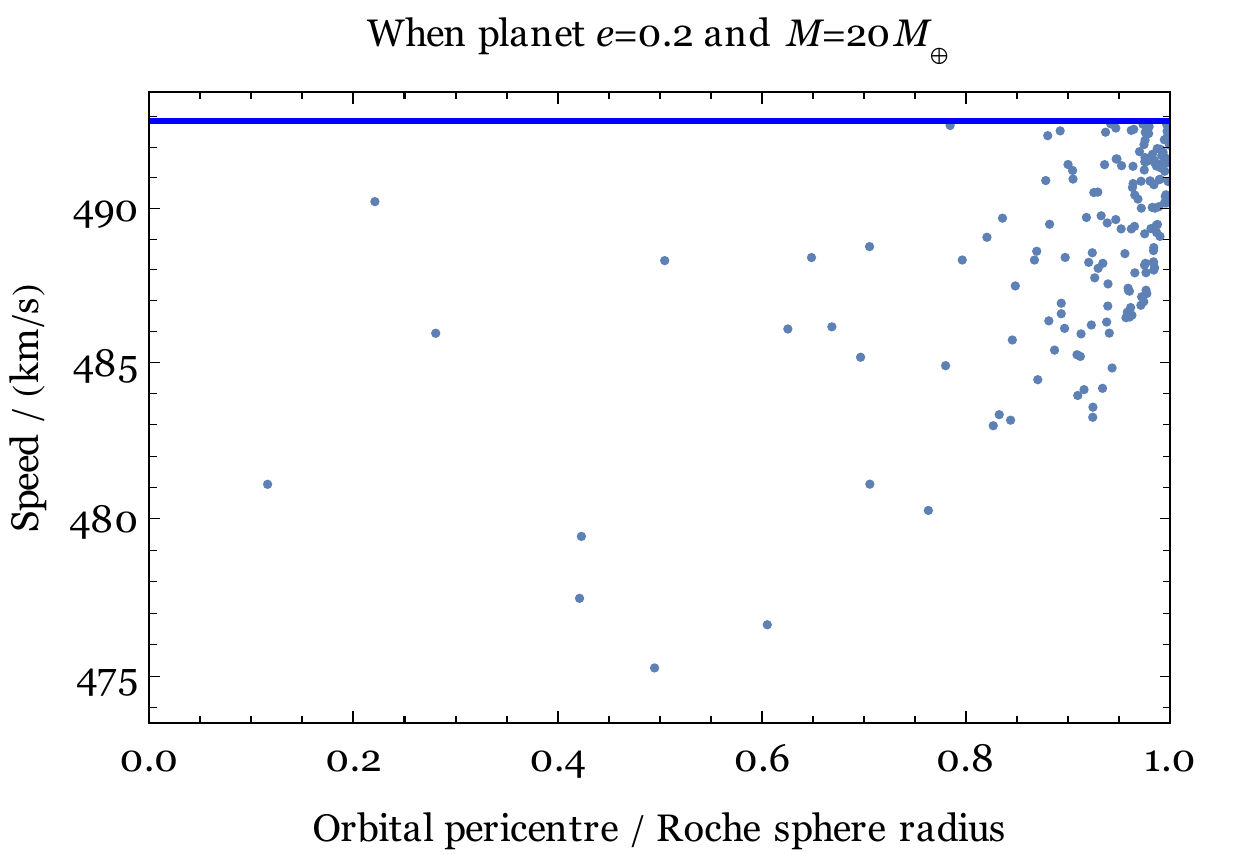}
\includegraphics[width=8.5cm]{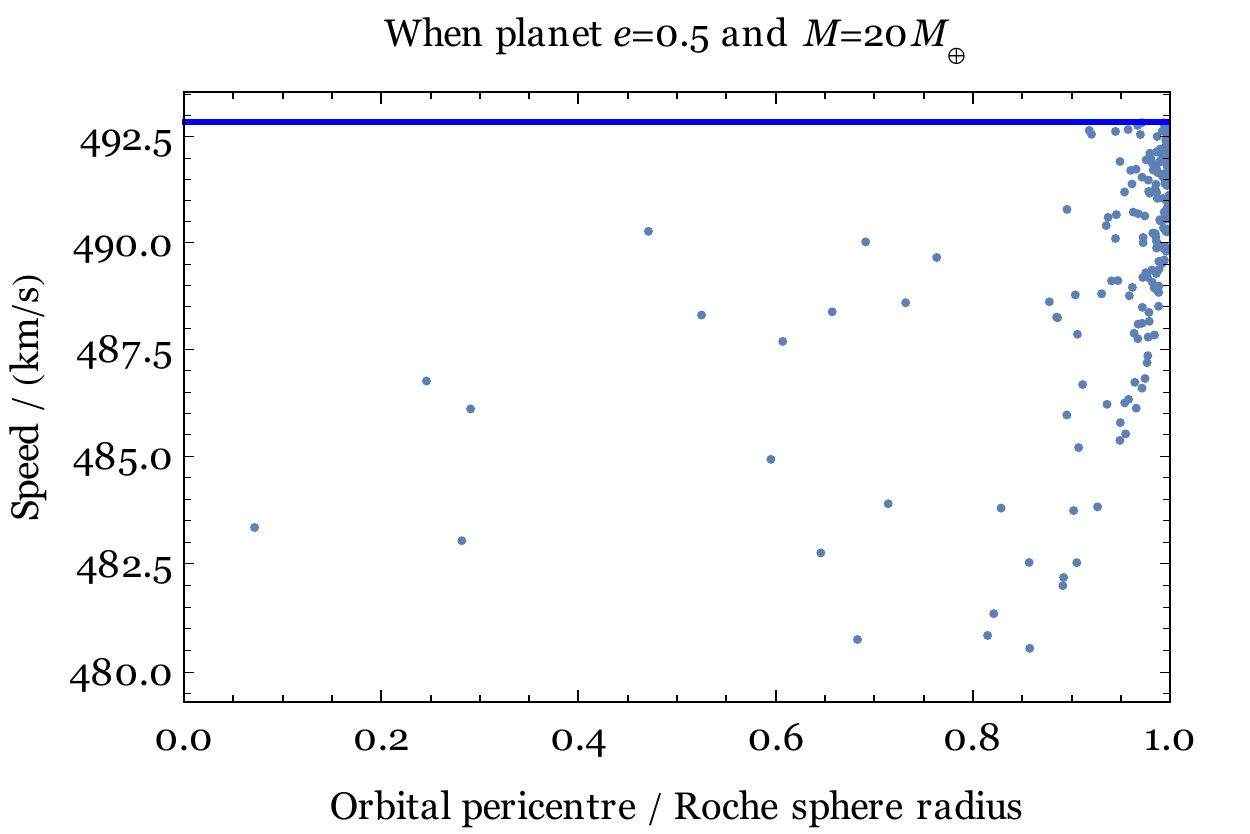}
}
\centerline{}
\centerline{
\includegraphics[width=8.5cm]{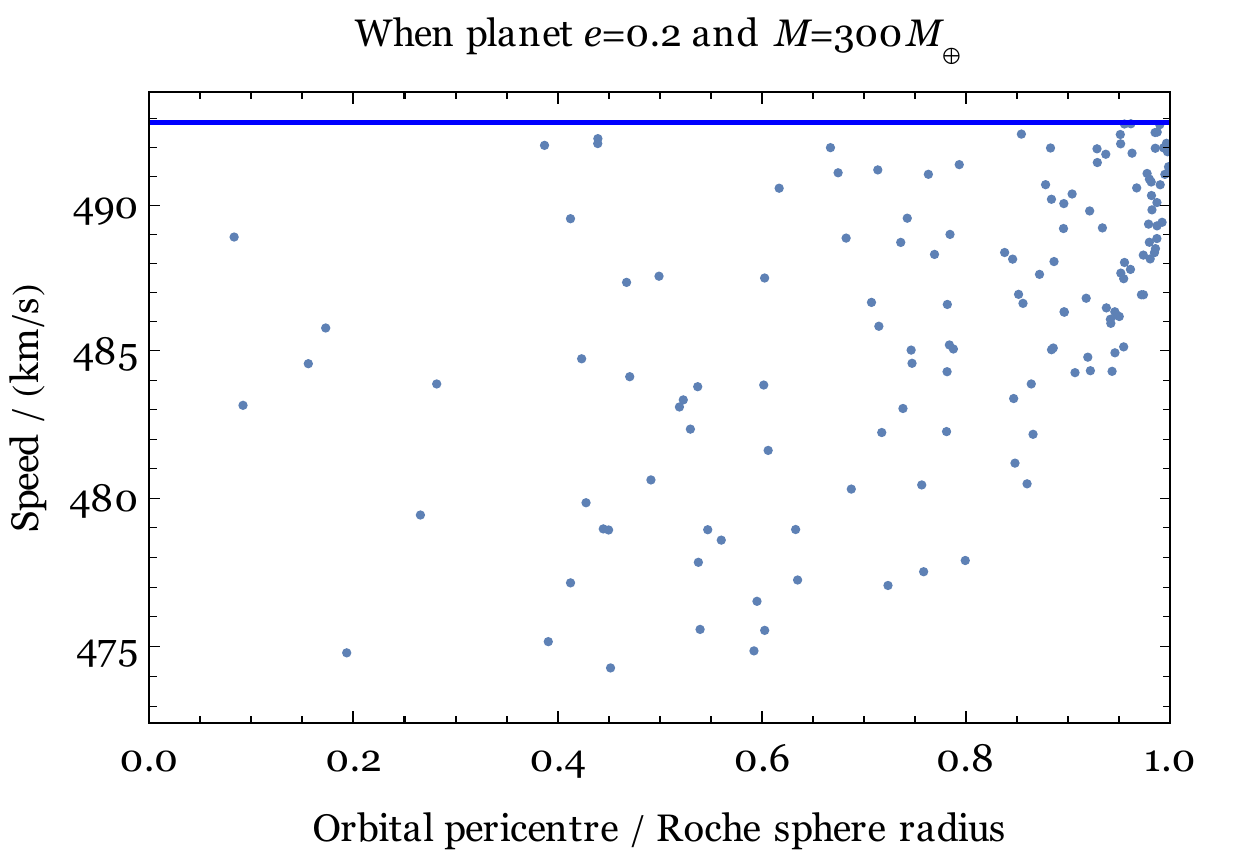}
\includegraphics[width=8.5cm]{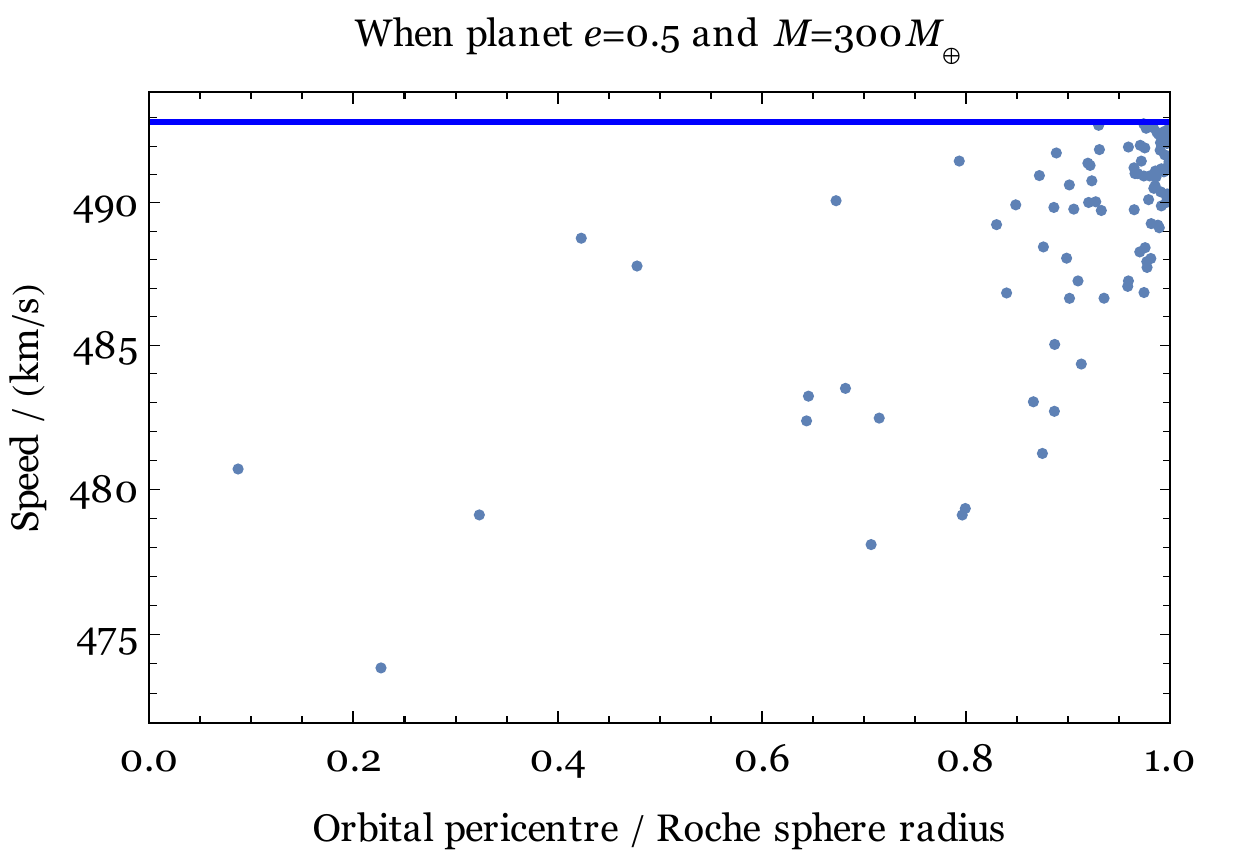}
}
\caption{
The entry speed into the white dwarf Roche sphere, as a function of osculating orbital pericentre immediately before engulfment.
}
\label{roche2}
\end{figure*}

\begin{figure*}
{\Huge How the osculating semimajor axis changes before engulfment}
\centerline{}
\centerline{
\includegraphics[width=8.5cm]{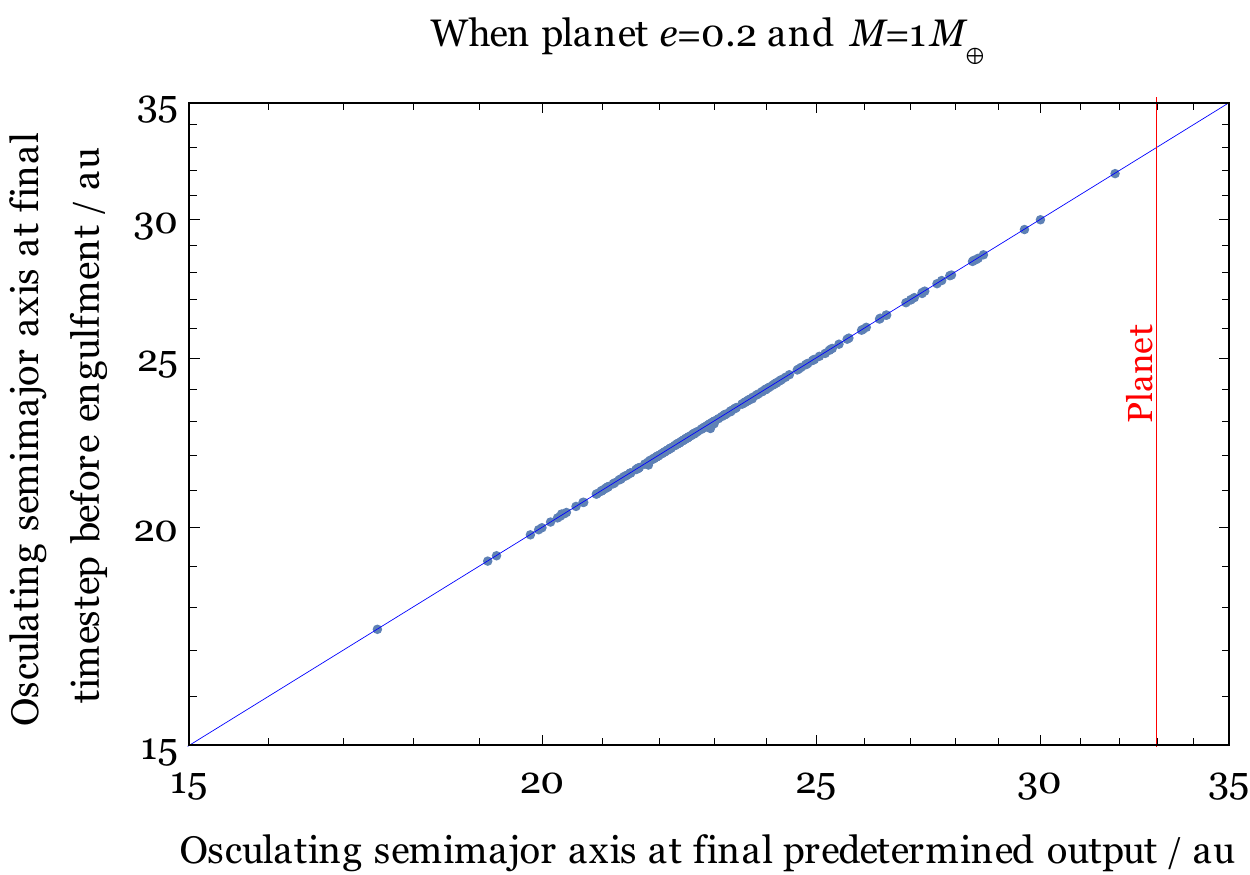}
\includegraphics[width=8.5cm]{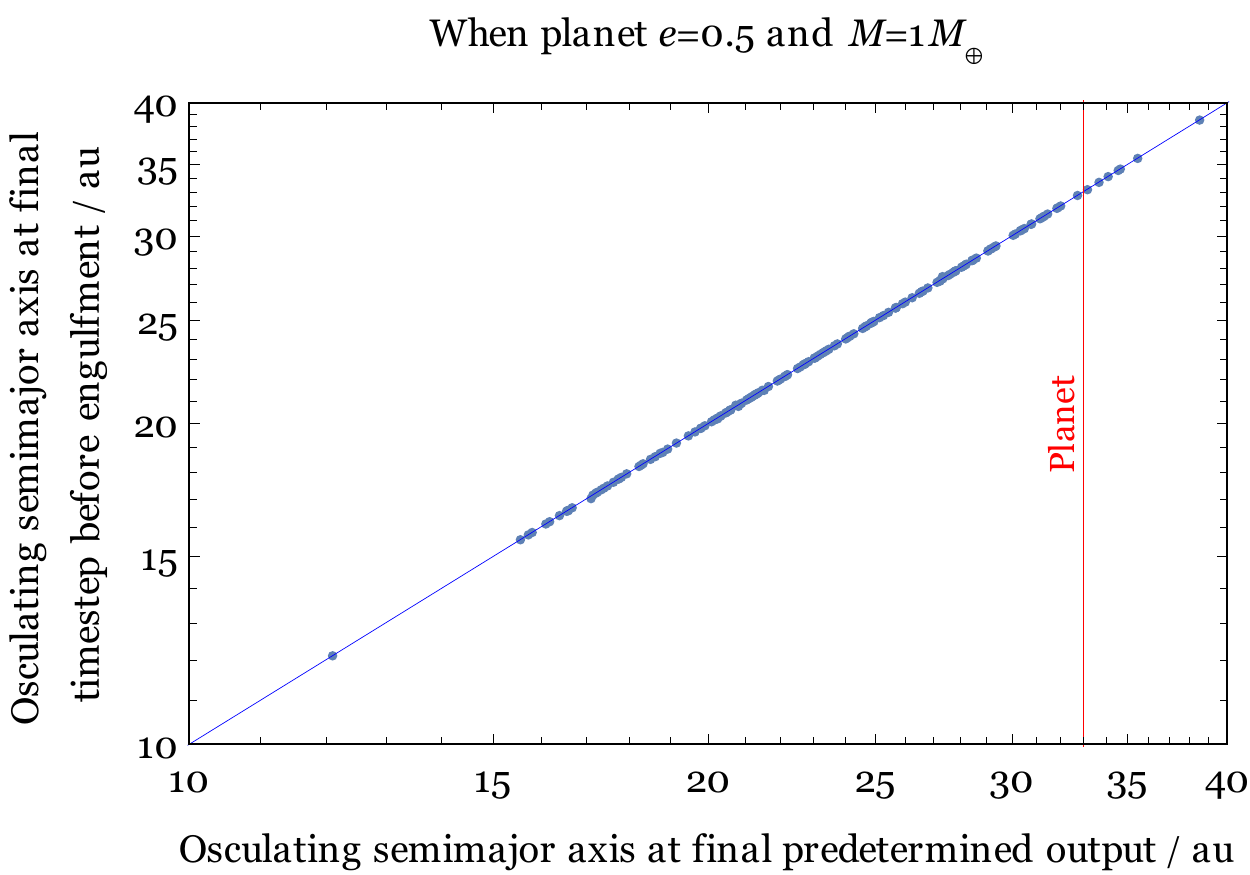}
}
\centerline{}
\centerline{
\includegraphics[width=8.5cm]{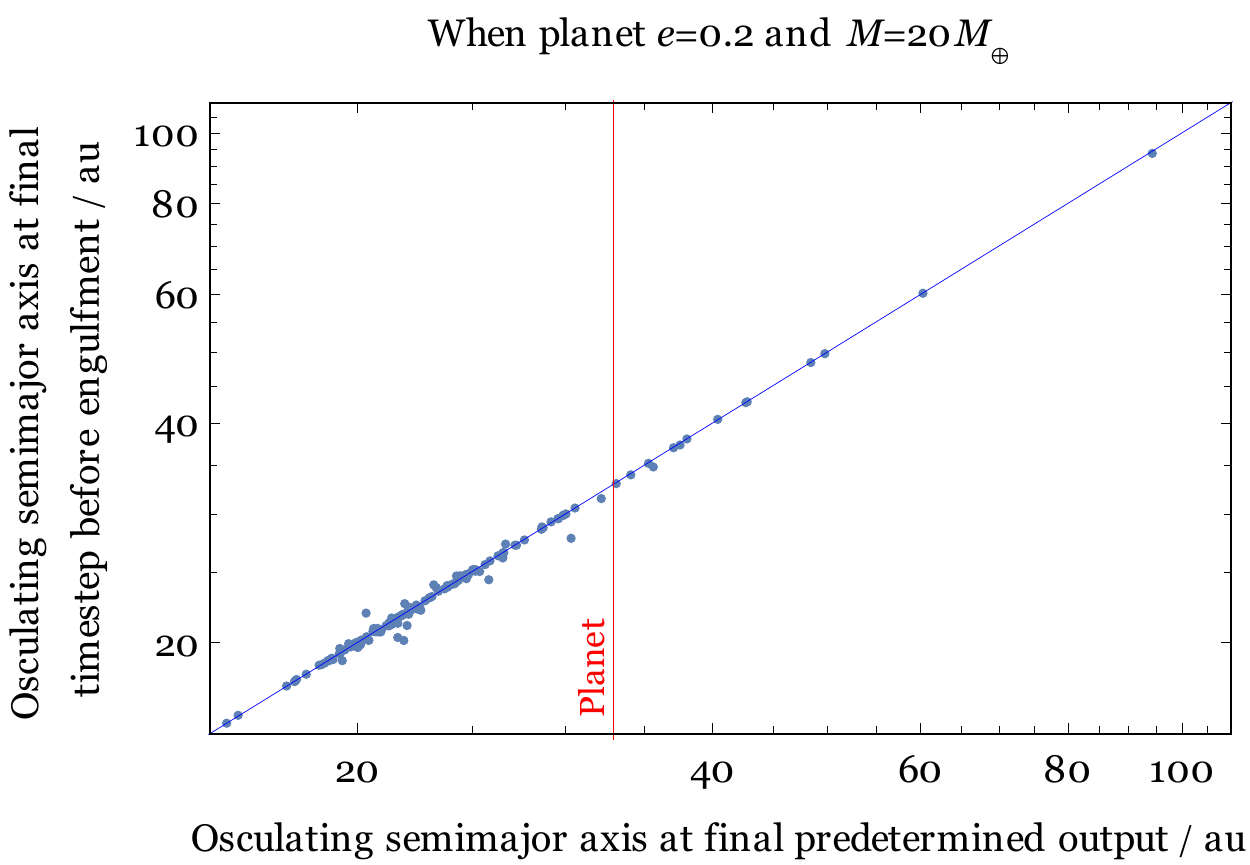}
\includegraphics[width=8.5cm]{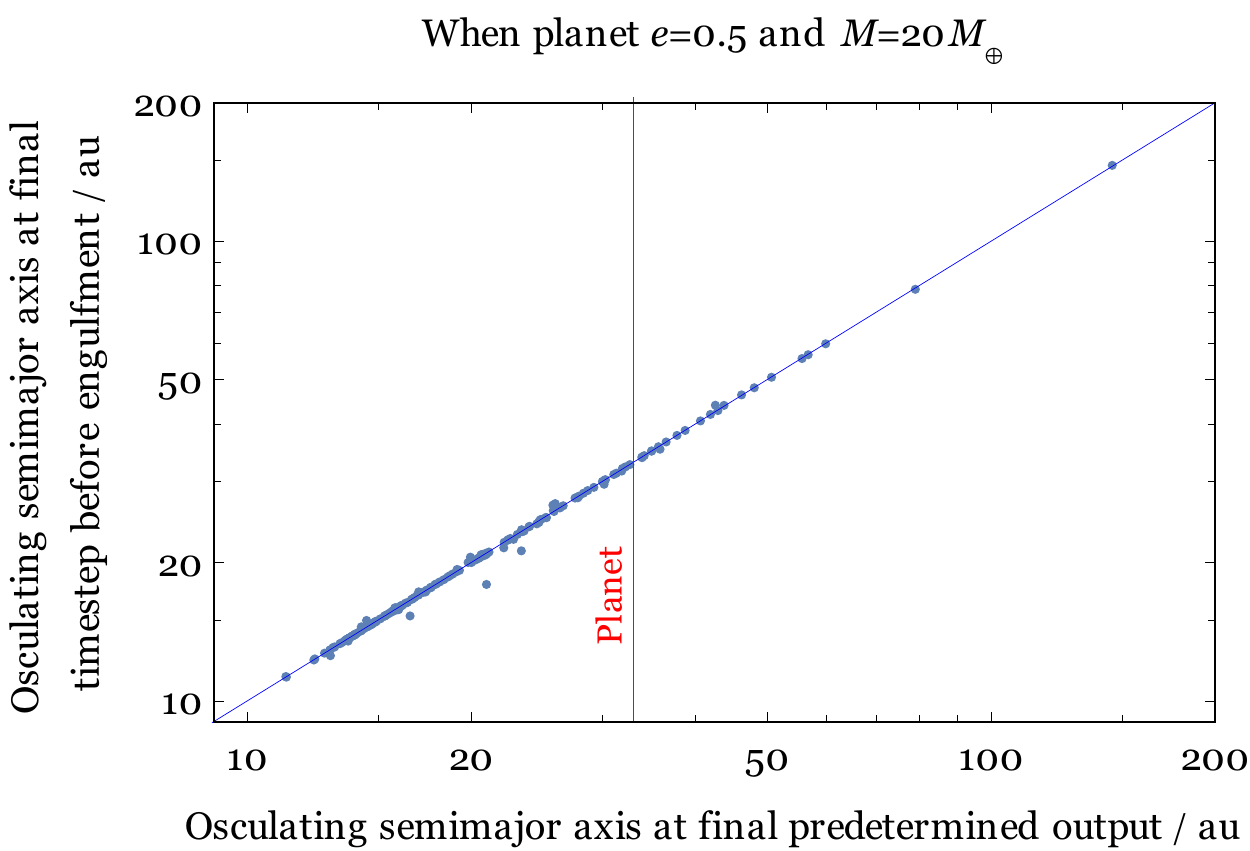}
}
\centerline{}
\centerline{
\includegraphics[width=8.5cm]{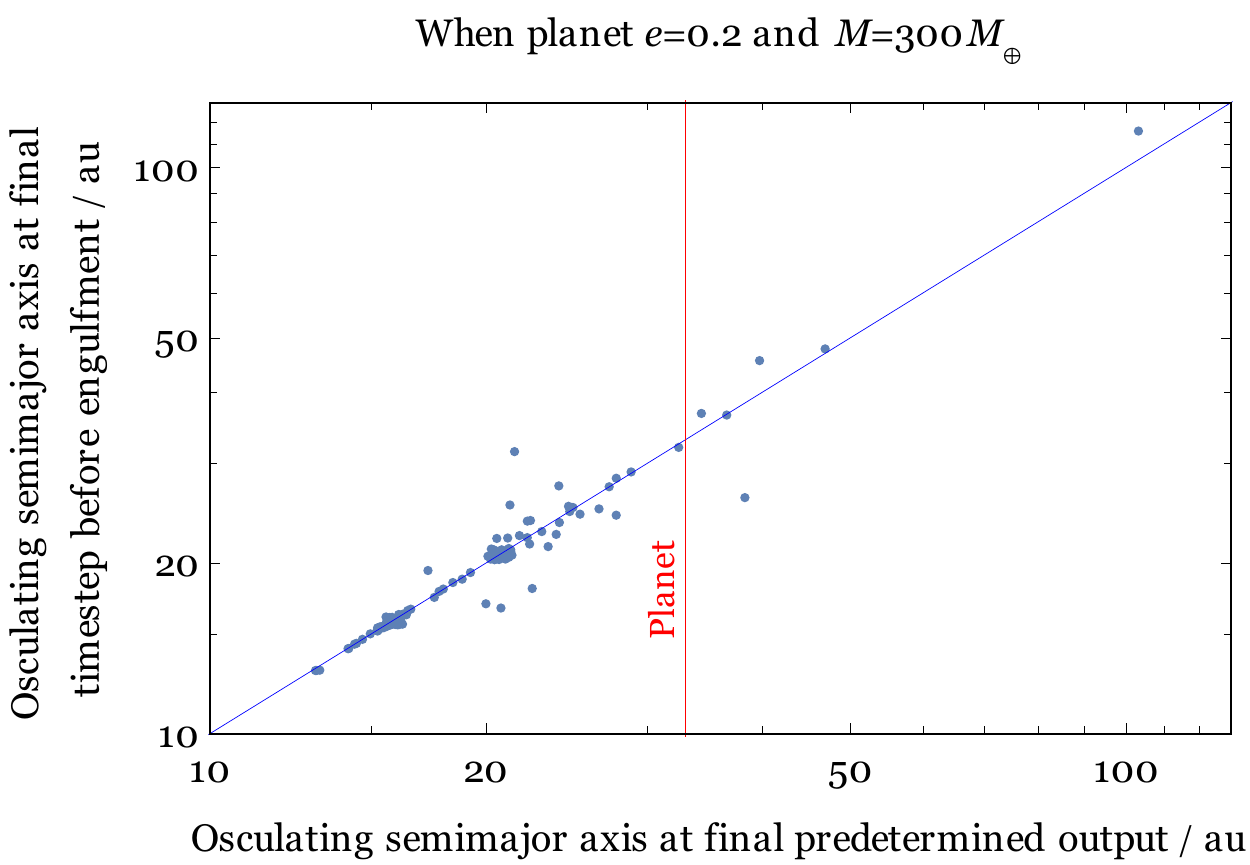}
\includegraphics[width=8.5cm]{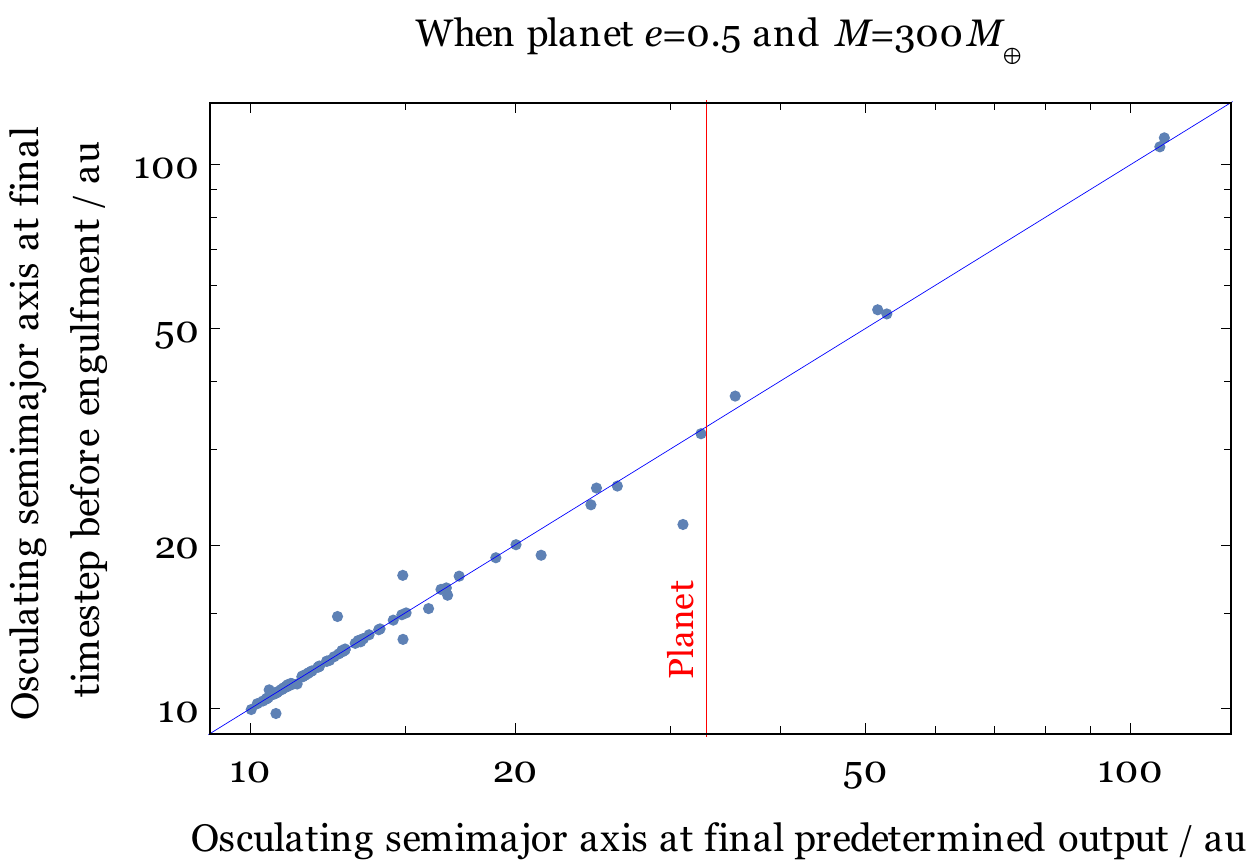}
}
\caption{
The osculating semimajor axis of the asteroid soon before engulfment, which determines the type of disc formed. The $x$-axis represents the last pre-determined simulation data output, which occurred at a frequency of about $1.33 \times 10^5$ yr from the start of the simulation. The $y$-axis represents the osculating semimajor axis at the last timestep before engulfment. This plot confirms the long-held notion that both $x$ and $y$ values should be similar, and reveals that scatter about this line is increased as the mass of the planet is increased.
}
\label{roche3}
\end{figure*}

\begin{figure*}
\includegraphics[width=14cm]{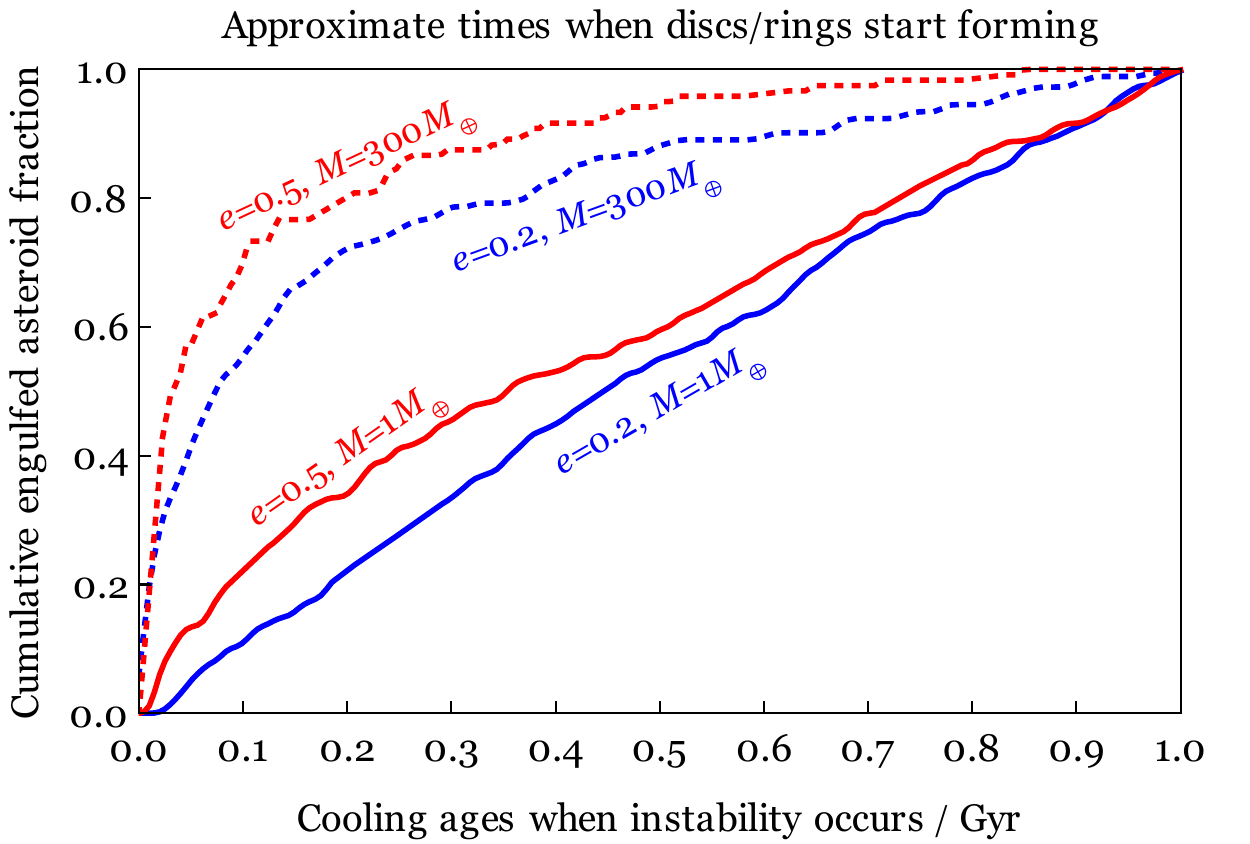}
\caption{
Instability times as a function of white dwarf cooling age for four cases. Shown are the cumulative distribution functions from the birth of the white dwarf to the end of the simulation, at a cooling age of 1 Gyr. The shape of these curves are a strong function of planetary mass, and suggest that for terrestrial planets, instabilities will continue more frequently at later cooling ages.
}
\label{Cooling}
\end{figure*}

\begin{figure*}
{\Huge Determining the morphology of discs/rings}
\centerline{}
\centerline{
\includegraphics[width=8.5cm]{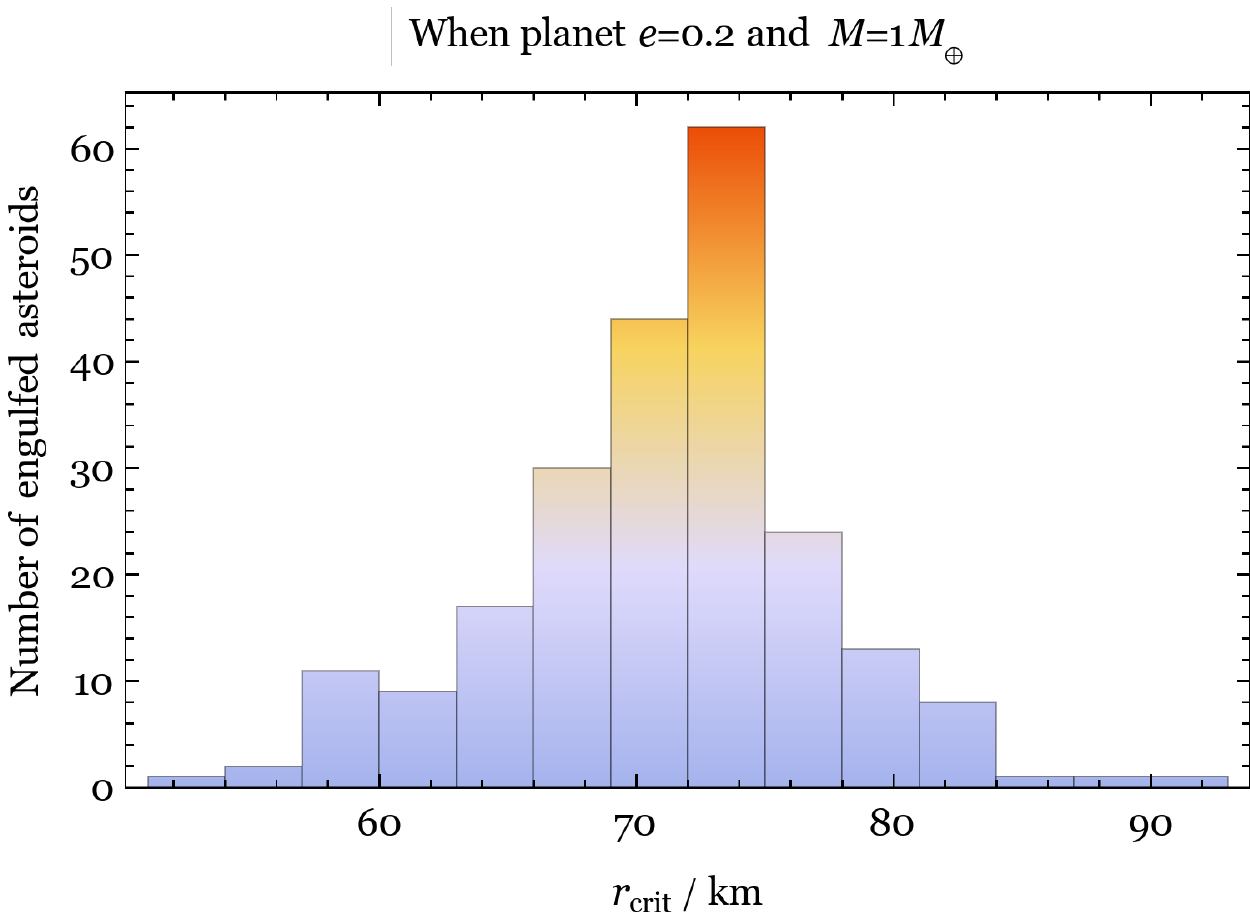}
\includegraphics[width=8.5cm]{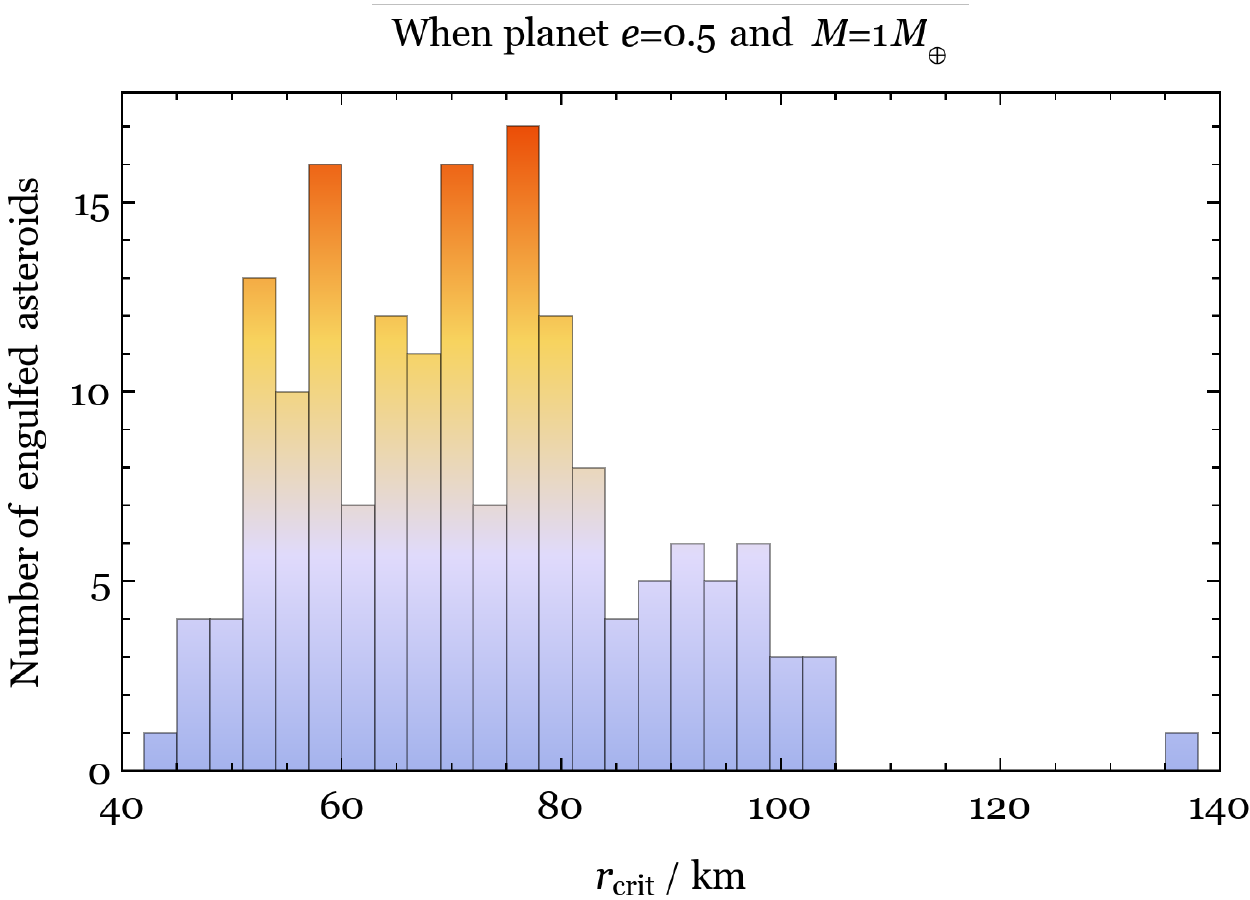}
}
\centerline{}
\centerline{
\includegraphics[width=8.5cm]{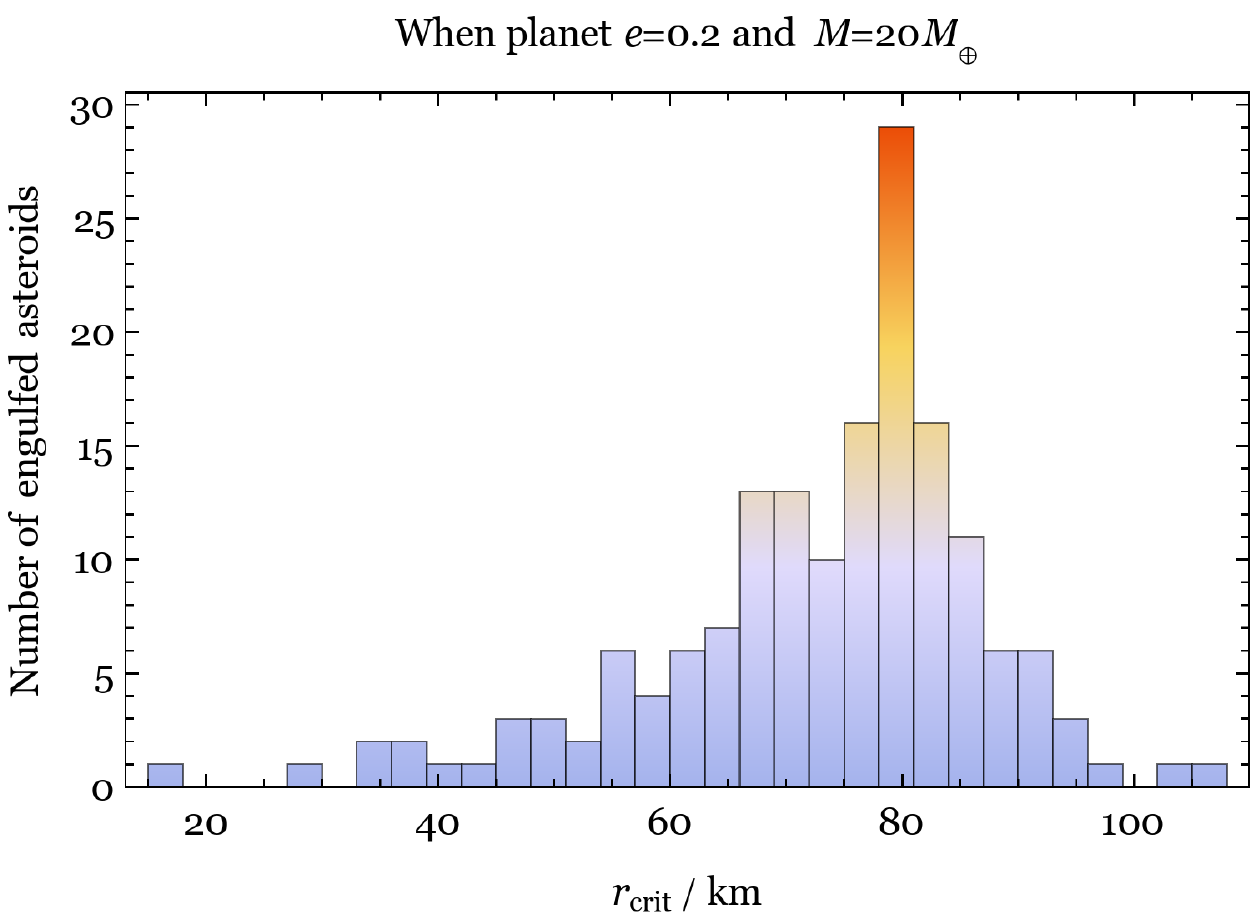}
\includegraphics[width=8.5cm]{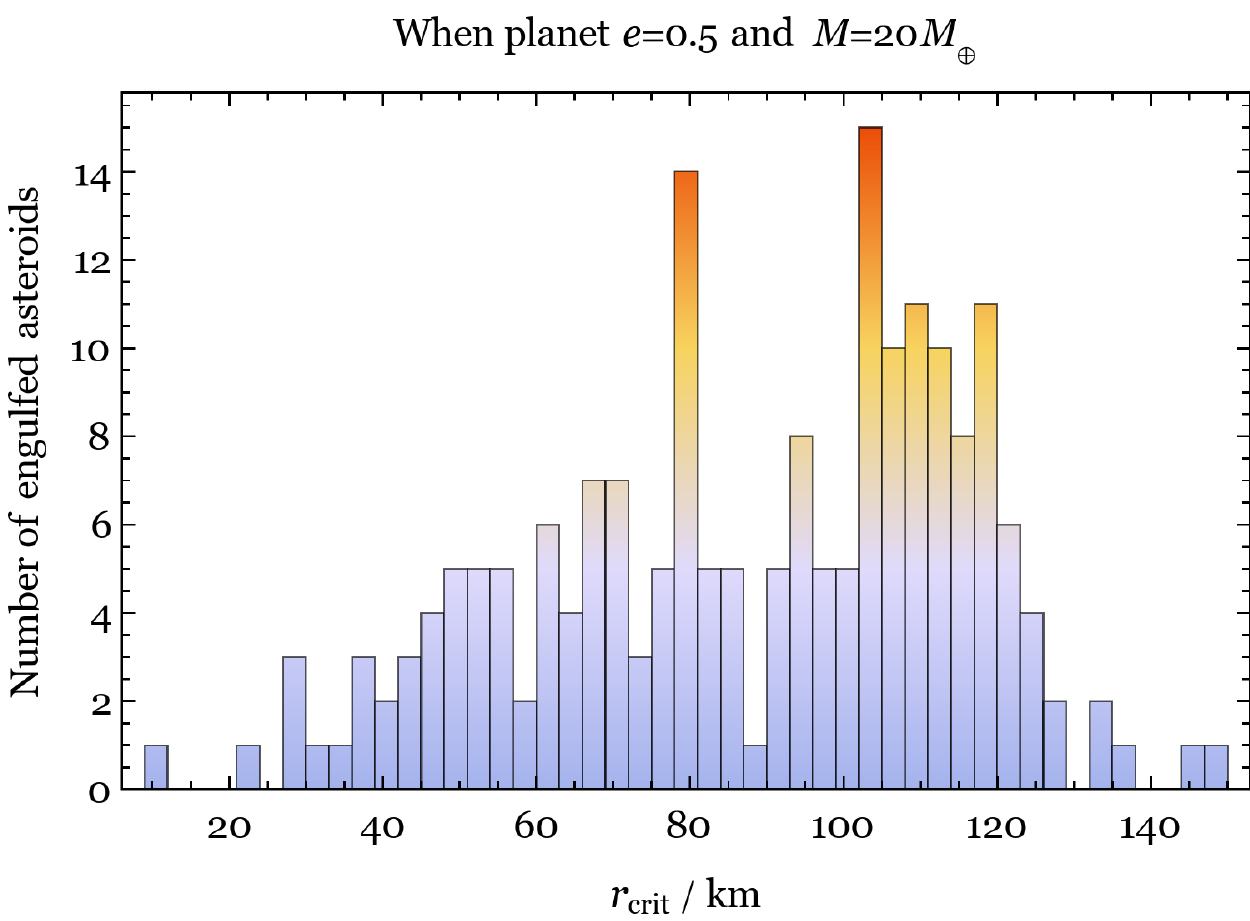}
}
\centerline{}
\centerline{
\includegraphics[width=8.5cm]{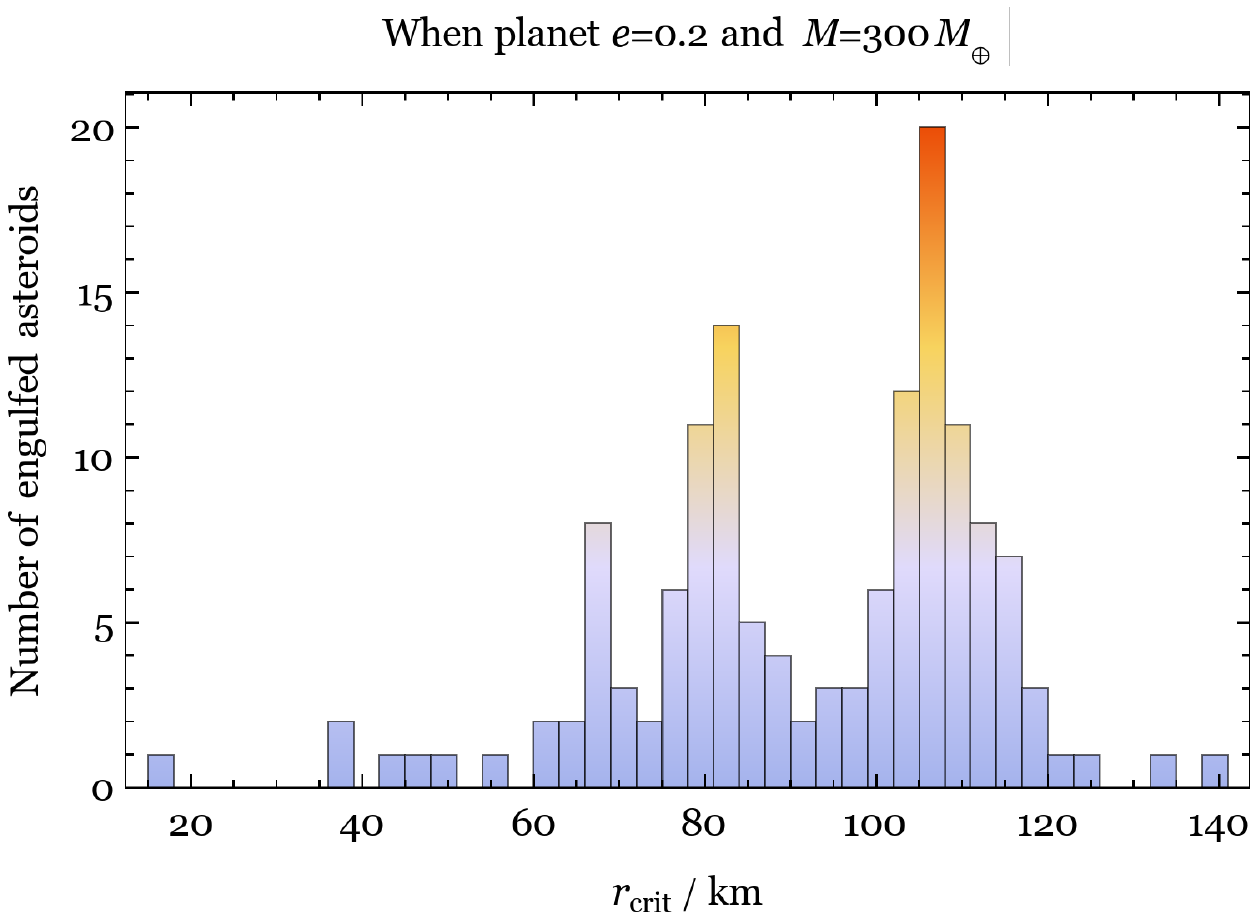}
\includegraphics[width=8.5cm]{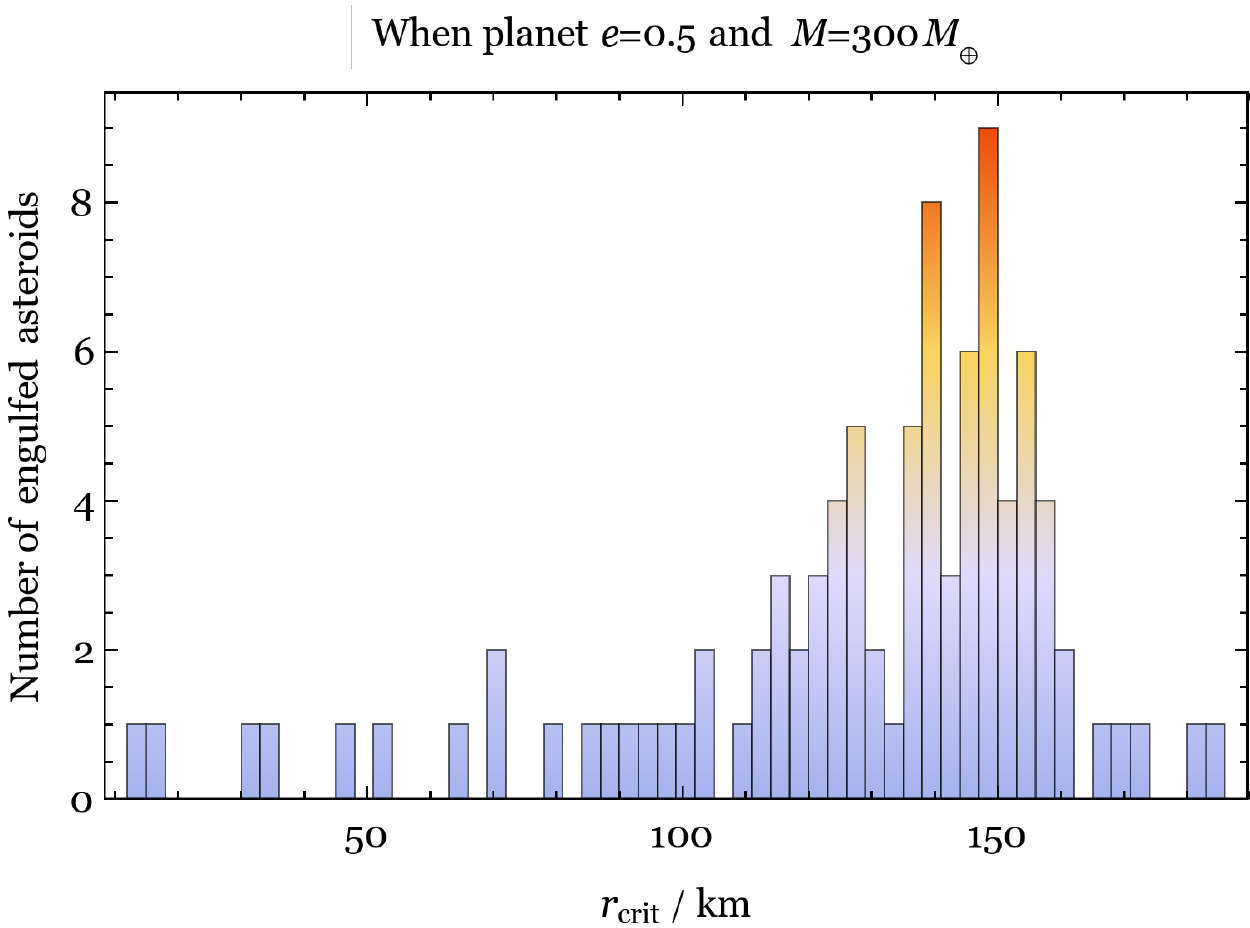}
}
\caption{
The value of $r_{\rm crit}$, which sets the asteroid radius boundary above which some of the disrupted fragments will be unbound. This value then hence determines the geometry of the resulting debris disc or ring \citep{malper2020a}.
}
\label{roche4}
\end{figure*}

\section{Numerical results}

Although the focus of this work is on the geometry of asteroid encounters with the white dwarf Roche sphere, our high-resolution simulations yield multiple dynamical results of potential interest. In this section, we present our results; their implications are discussed in Sections 4-5. {\rev We first provide a list of the percentages of asteroids which have crossed the white dwarf Roche sphere as a function of planet mass and eccentricity in Table \ref{stats}. This table illustrates that} the results of our simulations qualitatively differ depending on whether the planet's orbit is circular $(e=0.0)$ or eccentric $(e=0.2,0.5)$. {\rev We hence devote} separate subsections to each of these cases.

\begin{table}
	\centering
	\caption{{\rev The approximate percentage of asteroids in each simulation which both survived the giant branch phases of evolution and then crossed the white dwarf Roche sphere as a function of planet mass (rows) and eccentricity (columns).}}
	\label{stats}
	\begin{tabular}{ccccc} 
		\hline
		                                   & & $e=0.0$ & $e=0.2$ & $e=0.5$   \\
		\hline
		  $1M_{\oplus}$                   & &    & 3.3 & 2.2   \\
		  $5M_{\oplus}$                   & &    & 4.7 & 2.4   \\
		  $20M_{\oplus}$                  & &  0 & 2.3 & 2.6   \\
		  $100M_{\oplus}$                 & &    & 2.1 & 1.9   \\
		  $300M_{\oplus}$                 & &  0 & 1.8 & 1.2   \\ 
		\hline
	\end{tabular}
\end{table}

\subsection{The circular planet cases}

Circular planetary orbits fail to perturb asteroids into the white dwarf Roche sphere, a result known from both numerics \citep{frehan2014} and analytics \citep{antver2016,antver2019}. Here we performed two circular orbit simulations with planetary masses $M=(20M_{\oplus},300M_{\oplus})$, and also found that in no case was an asteroid perturbed to the white dwarf Roche sphere (Fig. \ref{e00}).

In Fig. \ref{e00}, nearly all instances of instability along the white dwarf phase are in the form of escape. For $M=20M_{\oplus}$, escape occurs predominately for asteroids whose initial apocentre exceeds that of the planet ({\rev purple} dashed line). The diagonal line featuring escape shows some instances of collisions occurring between the asteroid and planet. The escape speeds at the Hill surface along the white dwarf phase are all between about 0.01 km/s and 1 km/s, a range that helps distinguish the origin of this escape from other origins \citep{rafikov2018,malper2020b,pfaetal2021}. Instabilities occur throughout the simulations, and are likely to continue at older cooling ages.

The more massive planet ($M=300M_{\oplus}$) creates distinctly sharper features in these plots. First, on the left panel, the initial chaotic zone \citep{wisdom1980,muswya2012,decetal2013,petetal2018} has widened such that escape along the white dwarf phase occurs only for asteroids with an initial semimajor axis within about 9 au. Second, the inner escape boundary is still given by a diagonal line on which there are a few instances of collision with the planet. Third, the vertical lines of instabilities correspond to mean motion resonances, given by the dashed vertical lines. Of interesting note is the $3$:$2$ resonance, which features instabilities on both sides of the nominal resonance location. The right panel reveals clustering of the instability times due to the combination of this larger planet mass and stellar mass loss, although the escape speeds are comparable to the $M=20M_{\oplus}$ case.

\subsection{The eccentric planet cases}

When the planet is eccentric ($e=0.2,0.5$), the instability portraits along the white dwarf phase (Figs. \ref{inst1}-\ref{inst2}) now include asteroid encounters with the white dwarf Roche sphere (marked as ``Hit WD" on the legends). 

A comparison of the two figures reveals that the parameter space region in which these instabilities occur becomes more confined with increasing planet mass. In the $e=0.2$ case, this confinement is to mean motion resonances, whereas in the $e=0.5$ case, the confinement is primarily to the boundary of the sampled parameter space for the asteroid (the black solid curved line on the plots). Also, unlike in the $e=0.2$ cases, where escape is common, for each $e=0.5$ simulation, the predominant instability type is an encounter with the white dwarf Roche sphere. 

{\rev Another difference between the two figures is the prevalence of distinguishable resonant behaviour. Instability due to resonances can be more easily discerned for the highest planetary masses, and is not more prevalent for the $e=0.5$ cases because the resonances had already cleared away asteroids before the white dwarf phase. This last point is highlighted by the intermediate mass case of $20M_{\oplus}$, when the planet was not massive enough to generate such clearing. Table \ref{stats} demonstrates that overall, as the planet mass is increased, the number of particles crossing the Roche sphere decreases, reinforcing trends previously seen in the literature \citep{frehan2014,musetal2018}.}

Now we look more closely at these encounters, which is dependent on the geometry of the planetary orbit (Fig. \ref{PlanGeom}, following the standard convention from \citealt*{murder1999}). In Figs. \ref{e02P234} and \ref{e05P234}, we present projections of the Roche sphere overlaid with the final positions of asteroids before entering. 

{\rev The planet's orbital geometry and mass primarily dictate the anisotropy of the asteroid encounter geometry. As the planet mass increases, the anisotropy becomes less pronounced. The dependence on planet eccentricity is weaker, although the mass correlation is more pronounced in the $e=0.2$ case than the $e=0.5$ case.  For the highest mass sampled ($300M_{\oplus}$), encounters with the white dwarf Roche sphere in the $e=0.2$ case primarily arose from two mean motion resonances, whereas the encounters in the $M=300M_{\oplus}, e=0.5$ case did not. The reason is because in the $e=0.5$ case, these resonances had already cleared away asteroids during the giant branch phases.}

{\rev We note that the Cartesian points identified at the moments of encounter on Figs. \ref{e02P234} and \ref{e05P234} result from functional combinations of osculating orbital parameters. In the (standard) coordinate system that we adopted, an asteroid's $x$, $y$ and $z$ values are functions of its osculating semimajor axis, eccentricity, inclination, longitude of ascending node, argument of pericentre, and true anomaly \citep[see e.g.][]{murder1999}.}

Having established the physical location of entry, now we consider the velocity, both through the speed and the terminating osculating orbital elements (i.e. the elements recorded upon direct contact of the asteroid with the white dwarf Roche sphere). We present results for six cases of interest in Figs. \ref{roche1}-\ref{roche4}. 

In Fig. \ref{roche1}, the osculating orbital pericentres reveal how ``head-on" the asteroid encounter is with the Roche radius. In both $M=1M_{\oplus}$ cases, the anisotropy of the injections creates largely skirting encounters: here the smallest osculating pericentre distance is about $0.8R_{\odot}$. However, as the planet mass increases, the minimum preicentre distance shrinks, and in some cases can even reach values smaller than $0.1R_{\odot}$. While occasionally test particles of massive planets have more head-on trajectories, the typical outcome even for those planets is a skirting pericentre. Despite the high number of asteroids simulated, our sample size was not sufficiently high to determine statistics for collisional trajectories with the white dwarf photosphere, at about $0.01R_{\odot}$ \citep{broetal2017,mcdver2021}. 

Figure \ref{roche2} illustrates that the speed of entry is always confined to the range 474-493 km/s, which is just under the escape speed at the Roche sphere (493 km/s). Within this range, the speed of entry does not appear to correlate strongly with osculating pericentre distance or planetary masses or orbits.

The geometry of the disc which is formed from breakup is a function of the asteroid's osculating orbit immediately before engulfment into the Roche sphere. Hence, in Fig. \ref{roche3}, we report how the asteroid semimajor axis changes from its last recorded simulation value, which is up to about $1.33 \times 10^5$ yr before engulfment. All plots demonstrate that the terminal semimajor axis value is nearly equal to the pre-scattered value. However, the scatter about this alignment increases with planetary mass.

Finally, another dynamical result of interest is the instability time distributions in the simulations. These instability times are proxies for the start of the formation of the resulting debris disc or ring. Figure \ref{Cooling} presents the distributions for four limiting cases. The results are consistent with expectations that lower planetary masses can deliver material to the Roche sphere at later times.

\section{Implications for disc formation and evolution} \label{disc}

The entry geometry and velocity of asteroids have important implications for both disc formation (this Section) and accretion onto the white dwarf photosphere (Section 5). 

Our numerical simulations remove asteroids which enter the Roche sphere; this removal is unphysical. In reality, unless the asteroid directly collides with the white dwarf, which, according to Figs. \ref{roche1}-\ref{roche2}, occurs at just the $\sim$ 1 per cent level (see related discussions in Section 3, and also in \citealt*{wyaetal2014} and \citealt*{veretal2014c}), the asteroid must first tidally disrupt and form a debris disc. The latter then continues to undergo evolution.

\subsection{Initial debris formation}\label{SSS:formation}

Following a tidal disruption of the asteroid, analytical arguments in \cite{malper2020a} constrain the semimajor axes of the ensuing tidal fragments. They could occupy a range of semimajor axes based on the precise origin and size of the asteroid in consideration.

Because the range of initial asteroid semimajor axes that we consider in this paper is 3-10 au, orbital expansion dictates that their semimajor axes during the white dwarf phase are in the range of approximately 9-31 au (a 3.1 expansion factor) from mass loss alone. Kilometre-sized asteroids with a similar range of semimajor axes have always been assumed to tidally disrupt and form a ring with a similar semimajor axis. Such a ring would have a very tiny spread in the orbital energies of the fragments \citep{veretal2014c,malper2020a,nixetal2020}.

Increasing the size of the progenitor asteroid would generate a more dispersed disc of fragments. For a quantitative understanding, see Fig. 2 in \cite{malper2020a}. That investigation illustrates how below a critical radius, $r_{\rm crit}$ -- which depends on both the origin of the asteroid and its tidal breakup distance (see their equation 4) -- all of the tidal fragments would remain bound to the white dwarf. At exactly $r_{\rm crit}$, the innermost semimajor axis in the tidal stream would be halved (see their equation 3). Exceeding $r_{\rm crit}$ would lead to some fragments becoming unbound from the white dwarf. If the asteroid radius $R_{\rm ast}$ satisfies the condition $R_{\rm ast} \gg r_{\rm crit}$, then half of all the tidal fragments will be unbound, giving rise to a highly dispersed debris disc.

Kilometre-sized asteroids have always been assumed to satisfy $R_{\rm ast} \ll r_{\rm crit}$ when originating below a few dozen au, resulting in virtually no spread in orbital energies. The results in Fig. \ref{roche3} finally allow us to confirm this hypothesis. From the osculating terminal semimajor axis values in that figure, we can compute $r_{\rm crit}$, which we display in Fig. \ref{roche4}. The values of $r_{\rm crit} \approx 10-150$ km indicate that for bi-modal debris discs to form, the test particles would need to be large moons or dwarf planets. Such objects are rare in planetary systems characterized by a power law size distribution.

Figure \ref{roche4} indicates that the peak in $r_{\rm crit}$ values amongst the six distributions ranges between $80-150$ km, depending chiefly on the planet mass and to a lesser extent its eccentricity. These peaks suggest that small km-sized asteroids are indeed more likely to form rings or discs with little dispersion. Instead, for asteroid radii of tens of km, we would already expect significant dispersion in the disc. The initial eccentricity of the formed debris will always be extremely high and approaching unity. 

In what follows we briefly discuss how such fragments might subsequently evolve to circularize and form more compact central discs. Our simulations show that over a 1 Gyr period, asteroids are continuously injected into tidal crossing orbits. Each asteroid disrupts to form an eccentric disc of debris as described in Section \ref{SSS:pre-existing}.

\subsection{Interaction with pre-existing compact disc}\label{SSS:pre-existing}

Circularization can take place without a pre-existing compact disc in the vicinity of the white dwarf -- as will be discussed in the next subsections -- or indeed with the influence of such a disc, as will be discussed now. Previous studies by \cite{jura2008} and more recently \cite{ocolai2020} and \cite{maletal2021} considered the physical effects which a pre-existing compact disc near the white dwarf might have on the tidal fragments crossing it (in terms of fragment erosion or orbital circularization). The former two studies focused on gaseous compact discs while \cite{maletal2021} broadly discussed gaseous as well as dusty compact discs.

The study by \cite{maletal2021} identified two possibilities: (a) Suppose a massive asteroid forms an unusually massive compact disc around the white dwarf, and subsequently a typical-mass asteroid then crosses this disc. The interaction between the ensuing tidal stream and compact disc would lead to rapid and full circularization of the tidal fragments, and in particular faster circularization for the smaller fragments, for which drag-assisted circularization is more effective. In other words, the fragments become embedded in the compact disc, contributing their mass to that of the whole compact disc. The condition for this scenario to hold is that the compact disc mass must remain large between subsequent injections (i.e. avoiding full accretion onto the white dwarf). 

Our current medium- and high-mass planet simulations suggest that approximately $\sim 10^2$ ($\sim 1\%$) of the initially $10^4$ asteroids per simulation were injected into the Roche sphere over a period of 1 Gyr. If we scale up to an analogue asteroid belt containing $\sim 10^6$ asteroids over 1 km in size \citep{teddes2002}, then we generate approximately $10^4$ injections in 1 Gyr, or 0.1 Myr between injections. The empirical disc lifetime from \cite{giretal2012} is $10^4-10^6$ yr. This timescale comparison illustrates the plausiblity of massive compact discs persisting between injections, given our present scenario.

(b) In the more likely case that the debris disc is of comparable or smaller mass compared to the subsequent injected asteroid, then full circularization of the tidal stream is an impossibility. \cite{maletal2021} however showed that the outcome then would be the complete dispersal of the pre-existing compact disc, while the tidal stream undergoes partial circularization only. Even if the mass of the tidal stream exceeds the pre-existing compact disc mass by up 3-5 orders of magnitude (see Equation 8 in \citealt*{maletal2021}), significant partial circularization is still possible. We thus expect partial drag-assisted circularization to be a ubiquitous feature if even a low-mass pre-existing compact disc remains present between injections. Because the partial circularization is more effective for small tidal fragments, the reduction of the semimajor axes is greater for smaller fragments. Hence, partial circularization significantly shrinks the debris discs while also dispersing the orbital energy of the tidal fragments. We will later show that this fact can, in itself, significantly expedite the subsequent circularization by other mechanisms.

\subsection{Radiation effects}\label{SSS:radiation}

There are two potential radiation effects which may induce forces that would circularize and shrink an initially highly eccentric disc:

(a) The Yarkovsky effect is potentially capable of shrinking the tidal fragment orbits \citep{veretal2015b}. This effect is important for decimetre to 10 km sized fragments. According to various arguments in Section 2.2 of \cite{maletal2021}, the initial size of tidal fragments falls exactly within that range. However, better theoretical understanding of the seasonal Yarkovsky orbital shrinking effect in highly eccentric orbits is still required \citep{veretal2015a,veretal2019,veras2020}.

(b) Alternatively, many investigators \citep{bocraf2011,rafikov2011a,rafikov2011b,metetal2012,veretal2015b,veretal2021} also considered debris shrinkage through the drifting of small micron-to-cm sized dust by Poynting-Robertson drag.  However, whether this dust constitutes a significant mass fraction in the initial tidal stream is unclear. Nevertheless, collisional cascades can break initially large fragments to mere dust \citep{wyaetal2011} and then the tidal stream may shrink via Poynting-Robertson drag. We now discuss three ways with which to trigger such collisions among the fragments.

\subsection{Differential precession by general relativity}\label{SSS:GR}

Due to general relativity, the orbits of fragments inside the tidal stream must deviate from a perfect Keplerian one. Their pericentres precess, and with a rate proportional to the $-5/2$ power of the semimajor axis \citep{valetal2012}. Brouwers et al. (2021, in preparation) therefore recognize that if the initial eccentric debris disc is formed with some dispersion in the fragment orbital energies -- or else -- if fragments inside the tidal stream differentially alter their orbital energies through radiation effects (Section \ref{SSS:radiation}) or the effects of size-dependent partial circularization by a pre-existing compact disc (Section \ref{SSS:pre-existing}), then the inner fragments must precess more quickly than those on wider separations. Differential precession eventually leads to collisions among the fragments at pericentre. Gradually, the fragments are dissected to smaller bits and can be affected by Poynting-Robertson drag.

\subsection{Differential precession by a perturbing planet}\label{SSS:planet}

We postulate that a single perturbing planet has a rather similar effect to the one discussed in the previous paragraph. If we assume, by the same arguments, that the initial tidal debris disc forms dispersed, or evolves to become so, then we may look upon it as a collection of rings of various separations. If we treat each ring as a solid body rather than as a compilation of individual fragments, we can show that perturbations by the same planet which injected the test particle now causes subsequent precession to the ensuing tidal debris. The precession rate would be different in each ring, as follows.

Equations (174) and (175) of \cite{veras2014} show that, to leading order, the eccentricity and semimajor axis of the rings do not change. His Eq. (176), however, illustrates that the rate of precession goes as the $-3$ power of the semimajor axis (assuming a coplanar orbit for simplicity). Similar to the previous subsection, this dependence illustrates that differential precession would lead to collisions and in turn evolution by Poynting-Robertson drag, with two caveats. First, unlike in the previous subsection, this new mechanism is scenario-dependent and relies on the specific configuration of the system. It cannot work without an eccentric perturbing planet. Second, the secular treatment we invoked here considered rings as solids. We did not account for the possibility of the planet picking out individual fragments among the rings, which could actually conjure further collisions.

\subsection{Direct scattering by the planet}

In addition to the secular perturbations mentioned in the previous two sections, Brouwers et al. (2021, in preparation) and Li et al. (2021, in preparation) also recognize that certain portions (or even all) of the tidal stream continue to be directly scattered by the planet. In other words, if some tidal fragments closely approach the planet after the initial disruption, then they can be either ejected from the system or injected closer towards the white dwarf. In the latter case, fragments which were previously resistant to tidal disruption may newly disrupt upon close approach, or become more susceptible to sublimation or even directly collide with the white dwarf, expediting accretion.

We emphasize that fragments whose trajectories are exterior to the planet interaction zone are safe from direct planet scattering, but not from the secular perturbations by either the star or the planet, which were discussed previously.

\section{Implications for accretion onto the white dwarf}

Having described the formation and evolution of the debris as a function of geometry, we now provide an overview of the white dwarf physics that crucially determines the diffusion and detectability of debris accreted onto the photosphere at different locations.

\subsection{White dwarf characteristics}

White dwarfs have large surface gravities (somewhere in the range 10$^7$--10$^9$\,cm\,s$^{-2}$) which  
leads to gravitational settling of heavy elements towards the core \citep{schatzman1945,paqetal1986a,dupetal1992,koester2009}. This downward drift is driven by gravitational settling, thermal diffusion, radiative diffusion and diffusion driven by concentration gradients \citep{paqetal1986b,koester2009}. The result is that accreted metals sink below the photosphere into the deeper, non-visible layers. The timescale on which this sinking occurs depends strongly on the cooling age and composition of the white dwarf. As the white dwarf cools, two key processes delay the inevitable gravitational settling: radiative levitation and convection.

In young white dwarfs, heavy elements may be supported in the visible surface layers via radiative levitation. In H-atmosphere white dwarfs, levitation may play a role in observed surface abundances until the effective temperature decreases to below $\approx$20,000--30,000\,K \citep{chaetal1995}, or cooling ages of 10--50\,Myr \citep{fonetal2001}. In He-atmosphere white dwarfs, radiative levitation has little effect below effective temperatures of $\approx$40,000\,K, or cooling ages of over $\sim$4\,Myr. Without radiative levitation, the sinking timescales for H- and He-atmosphere white dwarfs is on the order of days and years, respectively \citep{koester2009,koeetal2020}. 

As the white dwarf cools, a superficial convection zone develops in the photospheric layers, allowing trace metals to be suspended in the visible layers. For He-atmosphere white dwarfs, the onset of convection occurs at cooling ages of 1--3\,Myr, or effective temperatures of $\approx$60,000--40,000\,K  \citep{fonetal2001,beretal2011,cuketal2018}. In H-atmosphere white dwarfs, which account for some 75--80\% of Milky Way white dwarfs \citep{torgar2016}, the onset of convection occurs at cooling ages of $\approx$100\,Myr, or an effective temperature of 18,000\,K \citep{cunetal2019}. As the white dwarf cools, the convection zone grows deeper, with its base reaching ever slower diffusing layers. This expansion increases the typical sinking timescales from days up to Myr \citep{cunetal2019,koeetal2020}.

\subsection{Linking to simulations}

\subsubsection{Homogenising timescales}

Accreted debris not only sinks radially towards the white dwarf core, but also spreads across the photospheric surface. \citet{cunetal2021} showed that homogenising debris across the surface of convective white dwarfs occurs on timescales on the order of 10$^1$--10$^5$\,yr. This range implies that surface abundances could exhibit some heterogeneity if metal accretion is highly localised, which will occur if this homogenising timescale is greater than the disc lifetime.

This homogenising timescale range is a subset of the disc lifetime range from theoretical constraints \citep {verhen2020}, which has no lower limit and an upper limit of $10^7$ yr. Empirical evidence from the currently observable sample \citep{giretal2012,cunetal2021} indicates a more restricted disc lifetime range of $\sim 10^4 - 10^6$ yr. Either way, the homogenising timescale may be longer than the disc lifetime.

\subsubsection{Convection zone timescales}

We can also compare disc lifetimes with the timescales of instability in Fig. \ref{Cooling} and with the timescales for white dwarfs to develop convection zones. Our simulations showed that for He-atmosphere white dwarfs and independent of initial planet mass, practically all engulfed asteroids arrive at the Roche sphere after the white dwarf has developed a convection zone, i.e., later than 3\,Myr. 

For H-atmosphere white dwarfs the picture is more nuanced. For our simulations which included a planet of mass 300\,$M_{\oplus}$, approximately 40--60 per cent of the engulfed asteroids arrived within the first 40\,Myr of white dwarf cooling, when radiative levitation was likely to still provide a dominant transport process in the surface layers. In this regime the presence of photospheric metals may be explained by primordial metals, or those accreted after the white dwarf forms \citep{baretal2014,koeetal2014}. 

For our simulations with a lower planet mass (1\,$M_{\oplus}$), approximately 80--90 per cent of {\rev tidally disrupted} asteroids encountered the Roche sphere after the white dwarf had developed a surface convection zone, i.e. after 100\,Myr of white dwarf cooling. In this scenario the origin of detected metals would be unambiguously planetary in nature.

\begin{figure}
\includegraphics[width=8.5cm]{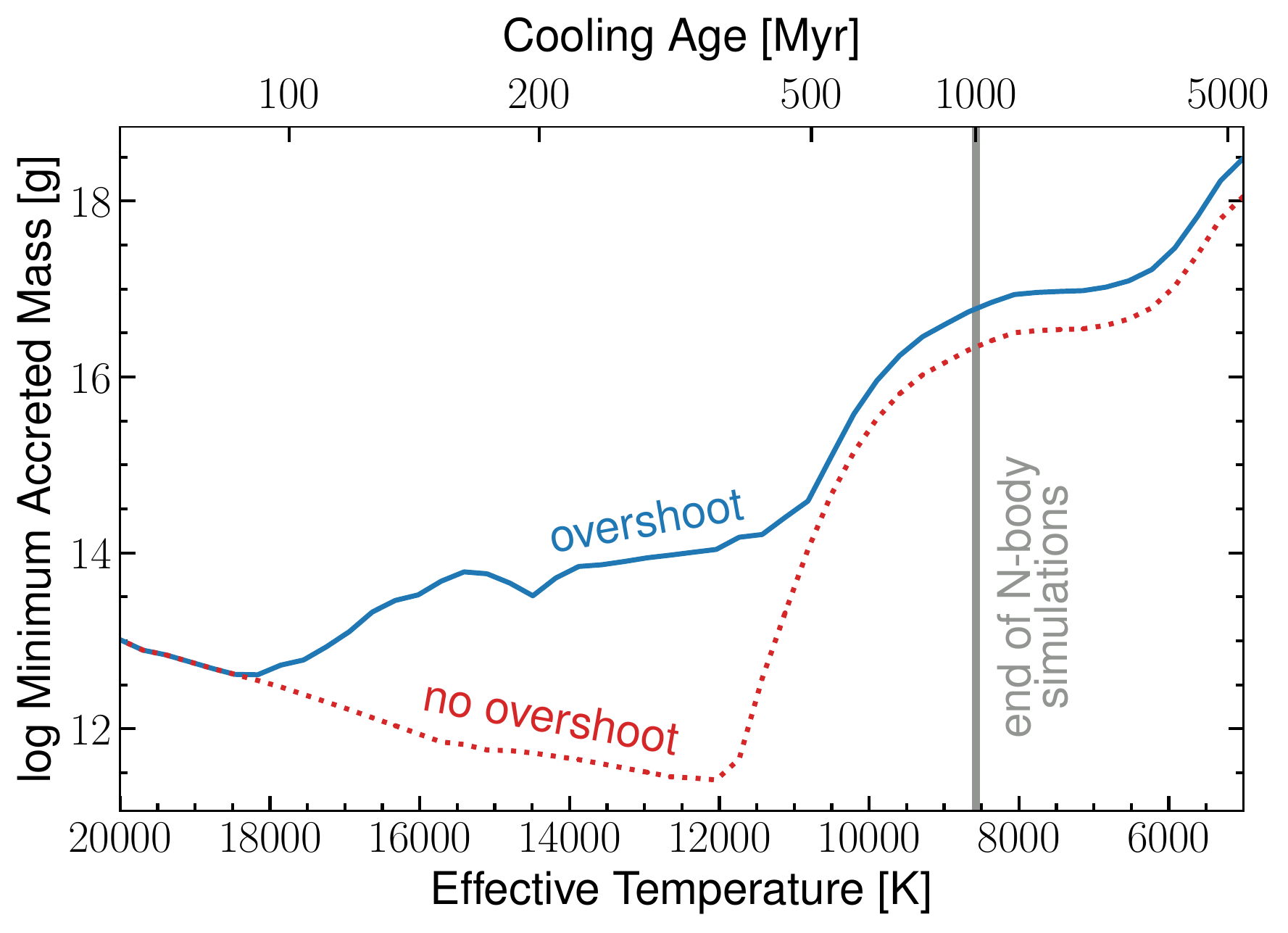}
\caption{
Minimum detectable debris mass within H-atmosphere white dwarfs as a function of age and stellar temperature. Plotted is the calcium detection threshold multiplied by convection zone mass with (blue) and without (red) the inclusion of convective overshoot, scaled to bulk Earth abundance. The calculation is for a white dwarf with a surface gravity of $\log g=8.0$. The calcium detection threshold, taken as a constant equivalent width of $EW=15$\,m\AA, is from Fig. 1 of Koester \& Wilken (2006). The convection zone masses are from Cunningham et al. (2019) and Koester et al. (2020) for the overshoot and no overshoot cases, respectively. The upper $x$-axis shows the corresponding cooling age for a $M=0.6 M_{\odot}$ white dwarf from the evolutionary models of Fontaine et al. (2001). The $N$-body simulations developed for this study cover the first Gyr of white dwarf cooling, illustrated by the grey vertical line in this plot. 
}
\label{diff}
\end{figure}

\subsubsection{Detectability}

The detectability of a single accretion event depends on, amongst other factors, the geometry of asteroid entry into the white dwarf Roche sphere and photosphere. For example, \citet{cunetal2021} performed a spectroscopic analysis of a metal-polluted DAZ white dwarf (SDSS J104341.53+085558.2) with a cooling age of $\approx$150\,Myr. They found that a sufficiently metallic spot covering just $\approx$10 per cent of the visible surface could be detectable in spectroscopic observations, although a more homogeneous distribution of surface metals better fit the observational data. Although observing such a feature is a possibility, the broad absorption line which would be produced by such a metallic spot has not yet been detected.

The ability to create such a spot depends on the disc evolution profile; probably the closer the asteroid pericentre is to the white dwarf photosphere, the sooner that the eventual debris will sublimate and accrete, although much more investigation about this transition is required. If the asteroid pericentre intersects with the white dwarf, then a direct impact will occur, but only if the asteroid is sufficiently strong and the white dwarf is sufficiently dim \citep{broetal2017,mcdver2021}.

Independent of whether the metals are all concentrated in a spot, or spread heterogeneously, or spread homogeneously, the minimum mass of metals required to be detectable can be estimated from the current minimum observed metal abundances. By using the measured calcium abundance, and assuming bulk Earth composition, one can compute the current minimum inferred accreted masses in H-atmosphere white dwarfs. These range from 10$^{13}$--10$^{18}$\,g \citep{giretal2012,koeetal2014,farihi2016}. The minimum detectable mass increases as the white dwarf cools due the deepening of the convection zone, a trend that we compute in Fig. \ref{diff}. Thus, whilst the longer sinking times of older white dwarfs increase the chances to detect a single accretion event, their larger convection zones require larger accreted masses to produce a detectable abundance.

\section{Summary}

We have performed a computationally-demanding suite of $N$-body simulations of one-planet systems with belts of test particles across the giant branch and white dwarf phases of evolution in order to investigate a variety of physics highlighted by the often-ignored entry geometry and velocity into the white dwarf Roche sphere. As a function of planet mass and eccentricity, we have characterized escape velocity (Fig. \ref{e00}), instability outcomes (Figs. \ref{e00}-\ref{inst2}), resonant behaviour (Fig. \ref{inst2}), entry geometry (Figs. \ref{e02P234}-\ref{roche1}), entry speed (Fig. \ref{roche2}), post-scattering semimajor axis values (Fig. \ref{roche3}), debris disc implications (Fig. \ref{roche4}) and instability times (Fig. \ref{Cooling}). We have also indicated the minimum detectable mass of our test particles (Fig. \ref{diff}). Overall, the anisotropy of minor body engulfment into the white dwarf Roche sphere may allow us to link detectable debris signatures with undetectable planetary architectures.

\section*{Acknowledgements}

{\rev We thank the reviewer for their helpful comments, which have improved the manuscript.}
We would like to thank the High Performance Computing Resources team at New York University Abu Dhabi and especially Jorge Naranjo for helping us with our numerical simulations. We also kindly thank Detlev Koester for providing the equivalent width data from Fig. 1 of \cite{koewil2006}. DV gratefully acknowledges the support of the STFC via an Ernest Rutherford Fellowship (grant ST/P003850/1), AJM acknowledges funding from the Swedish Research Council (starting grant 2017-04945), and TC has received funding from the European Research Council under the European Union's Horizon 2020 research and innovation programme n. 677706 (WD3D).

\section*{Data Availability}

The simulation inputs and results discussed in this paper are available upon reasonable request to the corresponding author.

\label{lastpage}
\end{document}